\documentclass[12pt]{article}
\usepackage{epsfig}
\usepackage{cite}

\def\beq{\begin{equation}}
\def\eeq{\end{equation}}
\def\beqa{\begin{eqnarray}}
\def\eeqa{\end{eqnarray}}

\newlength{\dinwidth} \newlength{\dinmargin}
\begin{document}

\begin{center}
{\Large \bf Soft-gluon corrections in top-quark production}
\end{center}
\vspace{2mm}
\begin{center}
{\large Nikolaos Kidonakis}\\
\vspace{2mm}
{\it Department of Physics, Kennesaw State University,\\
Kennesaw, GA 30144, USA}
\end{center}

\begin{abstract}
I review calculations of soft-gluon corrections for top-quark production in hadron collisions. I describe theoretical formalisms for their resummation and for finite-order expansions. I show that soft-gluon corrections are dominant for a large number of top-quark processes. I discuss top-antitop pair production as well as single-top production, including total cross sections and differential distributions, and compare with data from the LHC and the Tevatron. I also discuss top-quark production in association with charged Higgs bosons, $Z$ bosons, and other particles in models of new physics.
\end{abstract}

\section{Introduction}	

Top-quark physics is a central element in the exploration of particle physics
at hadron colliders. The top quark occupies a special place as the heaviest elementary particle to have been found, and the only quark that decays before it can hadronize. It was discovered via the top-antitop pair production process in proton-antiproton collisions at the Fermilab Tevatron by the CDF and D0 Collaborations in 1995 \cite{CDFttb95,D0ttb95}. Single-top production events were also first seen at the Tevatron \cite{D0singletop,CDFsingletop}. The top quark was later rediscovered at the LHC in $t{\bar t}$ \cite{CMSttb,ATLASttb} processes. The LHC now serves as a top-quark factory.

In this paper I review top-quark production in hadron colliders, focusing 
on higher-order corrections from soft-gluon resummation. 
There is a very long history of resummations for top-pair production, 
Refs. [7-67], for single-top production, Refs. [68-77], as well as for 
top-quark production in models of new physics, Refs. [78-84].
 
Next-to-leading order (NLO) \cite{NLOtt1,NLOtt2,NLOtt3} and next-to-next-to-leading order (NNLO) \cite{NNLOtt1,NNLOtt2,NNLOtt3,NNLOtt4} corrections have been available for top-pair production for some time. 
For single-top production, NLO corrections for the $t$ and $s$ channels \cite{BWHL} and for $tW$ production \cite{Zhu} are also known, while NNLO corrections \cite{NNLOtch,BGYZ,BGZ} have been calculated for the $t$-channel. Higher-order soft-gluon corrections can further improve the NLO and NNLO results.

In Section 2, I begin with a brief history of soft-gluon resummation, followed by 
a general discussion of higher-order soft-gluon 
corrections, factorization, renormalization-group evolution,  
resummation, and expansions at NLO, NNLO, and 
next-to-next-to-next-to-leading order (N$^3$LO).

In Section 3, I continue with a review of the cusp anomalous dimension and of soft anomalous dimension matrices for $t{\bar t}$ production through two loops. 
In Section 4, I provide results for the total $t{\bar t}$ cross sections, 
the top-quark transverse momentum, $p_ T$, distributions, and the top-quark 
rapidity distributions at the LHC and the Tevatron, as well as the forward-backward asymmetry at the Tevatron.

In Section 5, I discuss single-top production, including $t$-channel and 
$s$-channel production, and $tW$ production, and I present 
total cross sections and top-quark $p_T$ and rapidity distributions. 
In Section 6, I discuss top-quark production in association with a charged Higgs boson, and in association 
with gauge bosons via anomalous couplings in new-physics models.
I conclude with a summary in Section 7.

\section{Soft-gluon corrections}

Soft-gluon corrections arise from the emission of low-energy gluons, and they result from incomplete cancellations of infrared divergences between virtual diagrams and diagrams with real emission.

These corrections appear in the perturbative series as plus distributions involving logarithms of 
a variable that measures the kinematical distance from threshold. For the $n$th-order 
perturbative corrections, the leading logarithms are those with the highest power, $2n-1$; 
the next-to-leading logarithms have a power of $2n-2$; etc.  
The effects of soft-gluon corrections are particularly relevant near partonic threshold.
At partonic threshold there is no energy for additional radiation, but the top quark may have non-zero momentum and is not restricted to be produced at rest. Thus, partonic threshold is a more general concept than production or absolute threshold, where the top quark is produced at rest.

For top-antitop production, several threshold variables have been used for resummation. 
In single-particle-inclusive (1PI) kinematics, the partonic threshold variable is 
$s_4=s+t+u-\sum m^2$ where $s$, $t$, and $u$ are the standard kinematical variables and the sum is over the masses squared of all particles in the scattering. At partonic threshold, $s_4 \rightarrow 0$.
In pair-invariant-mass (PIM) kinematics, the partonic threshold variable  
is $1-z=1-M_{t{\bar t}}^2/s$, where $M_{t{\bar t}}$ is the invariant mass of the top-antitop pair; 
at partonic threshold $z \rightarrow 1$. 
In resummation using absolute threshold - a special limiting case of partonic 
threshold as we discussed above - the threshold variable is $\beta=\sqrt{1-4m_t^2/s}$, 
where $m_t$ is the top-quark mass; at absolute threshold, $\beta \rightarrow 0$.

Formalisms that use partonic threshold in 1PI or PIM kinematics involve a general resummation 
for double-differential distributions, from which single distributions or total cross sections 
can be derived by appropriate integrations. Formalisms that use absolute threshold are limited only to 
total cross sections.

Similarly, 1PI kinematics have been used in resummations for single-top production using the threshold variable $s_4$.

Soft-gluon corrections are dominant near threshold and usually even far from threshold. These corrections can be formally resummed to all orders in perturbative QCD. This resummation, i.e. exponentiation, follows from the factorization properties of the cross section and from the renormalization-group evolution of the functions in the factorized form. 

Leading-logarithm (LL) resummations and their finite-order expansions for top-antitop pair production were developed and used in various formalisms in Refs. \cite{LSvN1,LSvN2,NKJS,BC1,CMNT,BC2,CMNT2,BC3}. Next-to-leading-logarithm (NLL) resummations and their expansions were developed and used for double-differential cross sections in Refs. \cite{NKGS1,NKGS2,LOS,NKRV,NKrev,NK2000,KLMV,NKuni} using partonic threshold, and in Refs. \cite{BCMN1,BCMN2} for only total cross sections. Corrections beyond NLL in expansions were calculated in Refs. \cite{NK2000,NKuni,PRD68,NKrevmpla,NKnnnlo,MU,NKRV08}. Next-to-next-to-leading-logarithm (NNLL) resummations and expansions were developed in Refs. \cite{NK2loop,MSS,BN,BFS1,CMS,FNPY1,FNPY2,AFNPY2,BFS2,NKtt,AFNPY3,NKtopy}.

Leading logarithms can be resummed in terms of universal terms for the emission of collinear and soft gluons; these universal terms only depend on the identity of the incoming and outgoing partons. However, at NLL accuracy and beyond \cite{NKGS1,NKGS2} resummation involves the process-dependent color exchange in the hard-scattering process.

More explicitly, in 1PI kinematics the soft-gluon terms are of the form 
$[\ln^k(s_4/m_t^2)/s_4]_+$ 
where $k \le 2n-1$ for the $n$th-order perturbative corrections,  
and $s_4$ is the kinematical distance from partonic threshold described above.
We define
\beq
{\cal D}_k(s_4)\equiv\left[\frac{\ln^k(s_4/m_t^2)}{s_4}\right]_+
=\frac{\ln^k(s_4/m_t^2)}{s_4} \theta(s_4-\Delta)
+\frac{1}{k+1} \ln^{k+1}\left(\frac{\Delta}{m_t^2}\right) \delta(s_4) \, ,
\label{plus1PI}
\eeq
where $\Delta$ is a small parameter that separates
the hard-gluon, $s_4 > \Delta$, and soft-gluon, $s_4 < \Delta$, regions.

In PIM kinematics the soft-gluon terms are of the form $[\ln^k(1-z)/(1-z)]_+$, where again $k \le 2n-1$ for the $n$th-order perturbative corrections, and $1-z$ is the kinematical distance from partonic threshold described above.
We define
\beq
{\cal D}_k(z)\equiv\left[\frac{\ln^k(1-z)}{1-z}\right]_+
=\frac{\ln^k(1-z)}{1-z} \theta(1-z-\Delta)
+\frac{1}{k+1} \ln^{k+1}(\Delta) \, \delta(1-z) \, ,
\label{plusPIM}
\eeq
where, again, $\Delta$ is a small parameter that separates
the hard-gluon, $1-z > \Delta$, and soft-gluon, $1-z < \Delta$, regions.

\subsection{Factorization and Resummation}

We consider top-quark production in proton-proton collisions at the LHC,
\beq
p(p_A)+p(p_B) \rightarrow t(p_t)+X 
\eeq
and proton-antiproton collisions at the Tevatron
\beq
p(p_A)+{\bar p}(p_B) \rightarrow t(p_t)+X 
\eeq
where $t$ denotes the observed top quark, with $X$ all additional final-state particles. 
The partonic processes involved are of the form
\beq
f_{1}(p_1)\, + \, f_{2}\, (p_2) \rightarrow t(p_t)\, + \, X 
\eeq
where $f_1$ and $f_2$ are partons (quarks and/or gluons).
We define the usual kinematical variables $s=(p_1+p_2)^2$, $t=(p_1-p_t)^2$, $u=(p_2-p_t)^2$. We also define, as before, the 1PI partonic threshold variable $s_4=s+t+u-\sum m^2$. 

The factorized form of the (differential) cross section may be written in 1PI 
kinematics as 
\beq
d\sigma^{pp \rightarrow tX}=
\sum_{f_1,f_2} \; 
\int dx_1 \, dx_2 \,  \phi_{f_1/h_1}(x_1,\mu_F) \, 
\phi_{f_2/h_2}(x_2,\mu_F) \, 
{\hat \sigma}^{f_1 f_2 \rightarrow tX}(s_4,s,t,u,\mu_F,\mu_R)
\eeq
where $\mu_F$ is the factorization scale and $\mu_R$ is the renormalization scale. A similar expression applies to PIM kinematics. 
The parton distribution functions (pdf), $\phi$, describe the fraction of the proton momentum carried by the partons, with $x_1$ and $x_2$ the momentum fractions of partons $f_1$ and $f_2$ in protons $A$ and $B$ respectively.

Soft-gluon corrections in the partonic cross section, 
${\hat \sigma}_{f_1 f_2 \rightarrow tX}$,  
contribute terms with plus distributions of logarithms of $s_4$ or $1-z$, 
defined via their integral with the pdf. In 1PI kinematics
\beqa
\int_0^{s_{4 \, {\rm max}}} ds_4 \, \phi(s_4) \left[\frac{\ln^k(s_4/m_t^2)}
{s_4}\right]_{+} &\equiv&
\int_0^{s_{4\, {\rm max}}} ds_4 \frac{\ln^k(s_4/m_t^2)}{s_4} [\phi(s_4) - \phi(0)]
\nonumber \\ &&
{}+\frac{1}{k+1} \ln^{k+1}\left(\frac{s_{4\, {\rm max}}}{m_t^2}\right) \phi(0) \, , 
\eeqa
while in PIM kinematics
\beqa
\int_{z_{\rm min}}^1 dz \, \phi(z) \left[\frac{\ln^k(1-z)}{1-z}\right]_{+} &\equiv&
\int_{z_{\rm min}}^1 dz \frac{\ln^k(1-z)}{1-z} [\phi(z) - \phi(1)]
\nonumber \\ &&
{}+\frac{1}{k+1} \ln^{k+1}\left(1-z_{\rm min}\right) \phi(1) \, . 
\eeqa
These are equivalent definitions to Eqs. (\ref{plus1PI}) and (\ref{plusPIM}), respectively.

The resummation of soft-gluon contributions results from the factorization 
properties of the cross section in moment space.
Moments of the partonic cross section are defined by 
${\hat\sigma}(N)=\int (ds_4/s) \,  e^{-N s_4/s} \, {\hat\sigma}(s_4)$ in 1PI kinematics, and ${\hat\sigma}(N)=\int dz \, z^{N-1} \, {\hat\sigma}(z)$ in 
PIM kinematics.
Logarithms of $s_4$ or $1-z$ transform in moment space into logarithms of $N$,  
which exponentiate.

We write the moment-space partonic cross section, in $4-\epsilon$ dimensions, in factorized form, as
\beq
\sigma^{f_1 f_2 \rightarrow tX}(N, \epsilon)  
=\phi_{f_1/f_1}(N,\mu_F,\epsilon)\; 
\phi_{f_2/f_2}(N,\mu_F,\epsilon) \;
{\hat \sigma}^{f_1 f_2 \rightarrow tX}(N,\mu_F,\mu_R) 
\eeq
with  $\phi(N)=\int_0^1dx\; x^{N-1}\phi(x)$. 

We then refactorize the cross section \cite{NKGS1,NKGS2} as
\beqa
\sigma^{f_1 f_2\rightarrow tX}(N,\epsilon)
&=& 
H_{IL}^{f_1 f_2\rightarrow tX}\left(\alpha_s(\mu_R)\right) \, 
S_{LI}^{f_1 f_2 \rightarrow tX}\left(\frac{m_t}{N \mu_F},\alpha_s(\mu_R) \right)
\nonumber \\ && \hspace{-20mm} \times \;  
\prod  J_{\rm in} \left(N,\mu_F,\epsilon \right)
\prod J_{\rm out} \left(N,\mu_F,\epsilon \right)  \, ,
\eeqa
where $I$ and $L$ are color indices.

The $H_{IL}^{f_1 f_2\rightarrow tX}$ is the hard-scattering function 
that does not depend on $N$, and it describes contributions 
from the amplitude and the complex conjugate of the amplitude  
for the process. 
The soft-gluon function $S_{LI}^{f_1 f_2\rightarrow tX}$ describes 
the coupling of soft gluons to the partons in the scattering. 
The analytical form of $H_{IL}$ and $S_{LI}$ depends on the process, and in
general both functions are matrices in color space in the partonic scattering.
The $J_{\rm in}$ and $J_{\rm out}$ are jet functions that describe universal soft 
and collinear emission from incoming and outgoing massless partons.
 
The $N$-dependence of the soft matrix $S_{LI}$ can be resummed via 
renormalization group evolution \cite{NKGS1,NKGS2}.
We have 
\beq
S^b_{LI}=(Z_S^\dagger)_{LC}S_{CD}Z_{S,DI}
\eeq
where $S^b$ is the unrenormalized quantity 
and $Z_S$ is a matrix of renormalization constants.
Thus  $S_{LI}$ satisfies the renormalization group equation
\beq
\left(\mu \frac{\partial}{\partial \mu}
+\beta(g_s)\frac{\partial}{\partial g_s}\right)\,S_{LI}
=-(\Gamma^\dagger_S)_{LK}S_{KI}-S_{LK}(\Gamma_S)_{KI}
\eeq
where $g_s^2=4\pi\alpha_s$, and $\beta$ is the QCD beta function
\beq
\beta(\alpha_s) \equiv \frac{1}{2}\frac{d\ln \alpha_s}{d\ln \mu}
=-\sum_{n=0}^{\infty}\beta_n \left(\frac{\alpha_s}{4 \pi}\right)^{n+1}
\eeq
where $\beta_0=(11C_A-2n_f)/3$, $C_A=N_c$, with $N_c=3$ the number of colors, 
and  $n_f$ is the number of light quark flavors ($n_f=5$ for top production).

The soft anomalous dimension, $\Gamma_S$,
controls the evolution of the soft function $S$. 
We determine $\Gamma_S$ from the coefficients of the ultraviolet poles 
of eikonal diagrams. 
In dimensional regularization $Z_S$ has $1/\epsilon$ poles, and $\Gamma_S$
is given in terms of the residue of $Z_S$ \cite{NKGS1,NKGS2}.
In general, $\Gamma_S$ is a matrix in 
color space as well as a function of the kinematic variables $s$, $t$, $u$.

The resummed cross section in moment space is derived  
from the renormalization-group evolution of the functions in the 
factorized cross section, and can be written as
\beqa
{\hat{\sigma}}^{f_1 f_2\rightarrow tX}_{{\rm resum}}(N) &=&
\exp\left[ \sum_{i=1,2} E_i(N_i)\right] \, \exp\left[ \sum_j E'_j(N')\right]
\nonumber\\ && \times \,
\exp \left[\sum_{i=1,2} 2 \int_{\mu_F}^{\sqrt{s}} \frac{d\mu}{\mu}\;
\gamma_{i/i}\left({\tilde N}_i, \alpha_s(\mu)\right)\right] \;
\nonumber\\ && \times \,
{\rm tr} \left\{H^{f_1 f_2\rightarrow tX}\left(\alpha_s(\sqrt{s})\right)
\exp \left[\int_{\sqrt{s}}^{{\sqrt{s}}/{\tilde N'}}
\frac{d\mu}{\mu} \;
\Gamma_S^{\dagger \, f_1 f_2\rightarrow tX}\left(\alpha_s(\mu)\right)\right] \right.
\nonumber\\ && \left. \times \,
S^{f_1 f_2\rightarrow tX} \left(\alpha_s\left(\frac{\sqrt{s}}{\tilde N'}\right)
\right) \;
\exp \left[\int_{\sqrt{s}}^{{\sqrt{s}}/{\tilde N'}}
\frac{d\mu}{\mu}\; \Gamma_S^{f_1 f_2\rightarrow tX}
\left(\alpha_s(\mu)\right)\right] \right\} 
\nonumber \\
\label{resummed}
\eeqa
where the trace is taken of the product of the color-space hard and soft matrices, and the exponents of $\Gamma_S$.

The collinear-gluon and soft-gluon contributions from the initial-state partons  are resummed in the first exponential in Eq. (\ref{resummed}), with exponent 
\beq
E_i(N_i)=
\int^1_0 dz \frac{z^{N_i-1}-1}{1-z}\;
\left \{\int_1^{(1-z)^2} \frac{d\lambda}{\lambda}
A_i\left(\alpha_s(\lambda s)\right)
+D_i\left[\alpha_s((1-z)^2 s)\right]\right\}.
\eeq
Here we have defined $N_1=N(m_t^2-u)/m_t^2$ and $N_2=N(m_t^2-t)/m_t^2$. 
The perturbative series for $A_i$ is written as 
$A_i = \sum_{n=1}^{\infty}(\alpha_s/\pi)^n A_i^{(n)}$,
with $A_i^{(1)}=C_i$ \cite{GS87} where $C_i=C_F=(N_c^2-1)/(2N_c)$ 
for a quark or antiquark, while $C_i=C_A$ for a gluon; 
and $A_i^{(2)}=C_i K/2$ \cite{CT89} where 
$K= C_A\; ( 67/18-\zeta_2 ) - 5n_f/9$ \cite{KT82},  
with $\zeta_2=\pi^2/6$. 
Also the perturbative series for $D_i$ is written as $D_i=\sum_{n=1}^{\infty}(\alpha_s/\pi)^n D_i^{(n)}$, with $D_i^{(1)}=0$ in Feynman gauge.

The collinear-gluon and soft-gluon contributions from final-state massless quarks and/or gluons are resummed in the second exponential, with exponent  
\beqa
E_j'(N') &=& 
\int^1_0 dz \frac{z^{N'-1}-1}{1-z}
\left \{\int^{1-z}_{(1-z)^2} \frac{d\lambda}{\lambda}
A_j \left(\alpha_s\left(\lambda s\right)\right)
+B_j \left[\alpha_s((1-z)s)\right] \right.
\nonumber \\ && \hspace{28mm} \left.
+D_j \left[\alpha_s((1-z)^2 s)\right]\right\} 
\eeqa
where we have defined $N'=N s/m_t^2$. 
The perturbative series for $B_j$ is written as  
$B_j=\sum_{n=1}^{\infty}(\alpha_s/\pi)^n B_j^{(n)}$, 
with $B_q^{(1)}=-3C_F/4$ for a quark or antiquark, and $B_g^{(1)}=-\beta_0/4$ 
for a gluon \cite{GS87,CT89}.
Note that this exponent is not needed in $t{\bar t}$ or $tW$ production 
but it is used in $s$-channel and $t$-channel single-top production.

The factorization-scale dependence in the third exponential is given in 
moment space by the anomalous dimension of $\phi_{i/i}$, 
which is $\gamma_{i/i}=-A_i \ln {\tilde N}_i +\gamma_i$, 
where ${\tilde N}_i=N_i e^{\gamma_E}$ with $\gamma_E$ the 
Euler constant, and 
$\gamma_i=\sum_{n=1}^{\infty}(\alpha_s/\pi)^n \gamma_i^{(n)}$
with  $\gamma_q^{(1)}=3C_F/4$, $\gamma_g^{(1)}=\beta_0/4$.

The perturbative series for the hard function $H$ and 
the soft function $S$ are written as   
$H=\alpha_s^{d_{\alpha_s}} \sum_{n=0}^{\infty}(\alpha_s/\pi)^n H^{(n)}$
and
$S= \sum_{n=0}^{\infty}(\alpha_s/\pi)^n S^{(n)}$, 
respectively, where $d_{\alpha_s}$ denotes the power of $\alpha_s$ in the  
leading-order (LO) cross section.
The LO cross section for each partonic process is given by the trace of 
the product of the lowest-order hard and soft matrices: 
$\sigma^B=\alpha_s^{d_{\alpha_s}}{\rm tr}[H^{(0)}S^{(0)}]$.

Noncollinear soft-gluon emission is described by the soft anomalous dimension 
$\Gamma_S$, with perturbative expansion 
\beq
\Gamma_S=\sum_{n=1}^{\infty}\left(\frac{\alpha_s}{\pi}\right)^n \Gamma_S^{(n)}
\eeq
The first term in the series, $\Gamma_S^{(1)}$, requires one-loop calculations 
and is needed for NLL resummation, while
$\Gamma_S^{(2)}$ requires two-loop calculations
and is needed for NNLL resummation.

\subsection{Methods and prescriptions}

There are numerous and very substantive differences between the various resummation approaches in the literature, which have been detailed previously in Refs. \cite{NKBP,NKhq13}. Some formalisms have been developed to do the resummation only for the total cross section while others are for the double-differential cross section; some formalisms use moment-space perturbative QCD while others use Soft-Collinear Effective Theory (SCET). The treatment of subleading logarithms in different formalisms and the approach to deal with infrared divergences via prescriptions or finite-order expansions lead to large differences in numerical results.

The more general double-differential approach to resummation can be expressed in 1PI kinematics for the differential cross section in top-quark transverse momentum and rapidity, $d\sigma/dp_T dy$, where the soft limit is $s_4 \rightarrow 0$, or in PIM kinematics for the differential cross section in top-pair invariant mass and scattering angle, $d\sigma/dM_{t \bar t} \, d\cos\theta$, where the soft limit is $z \rightarrow 1$. Such double-differential approaches have been developed in moment-space in QCD, Refs. \cite{NKGS1,NKGS2,LOS,NKrev,NK2000,KLMV,NKuni,PRD68,NKrevmpla,NKnnnlo,NKtt,NKhq13,N3LO1,N3LO2}, as well as in SCET, Refs. \cite{FNPY1,FNPY2,AFNPY2,AFNPY3}. Double-differential distributions, single-differential distributions, and total cross sections can all be derived via the above formalisms. Resummations that are developed for the total cross section only refer to production (or absolute) threshold and resum logarithms of $\beta = \sqrt{1-4m_t^2/s}$, Refs. \cite{BCMN1,MU,BFS1,ALLMUW,BFS2,CCMMN}. The soft limit here is the production threshold limit $\beta \rightarrow 0$ (where the top quark velocities are zero), which is a special case of the more general partonic threshold.

The resummed cross section encounters infrared divergences that require a prescription to be dealt with, 
and the choice of prescription is to an extent arbitrary. 
Moreover, the numerical results depend greatly on the prescription, and differences between prescriptions 
are typically larger than corrections beyond NNLO. 
In order to avoid arbitrary prescription dependence, 
an excellent approach is to expand the resummed cross section to a fixed order, usually NNLO or N$^3$LO; this gives better control of subleading terms.

The relative size of soft-gluon corrections has been argued about since the 1990's. The LL results in Refs. \cite{LSvN1,LSvN2,NKJS} and \cite{BC1,BC2,BC3} argued for large effects. The LL results in \cite{CMNT,CMNT2}, using the minimal prescription, argued for tiny corrections, one or two orders of magnitude smaller. Similarly at NLL, the results in Ref. \cite{BCMN1}, which were based on the minimal prescription, were much smaller than the results in Refs. \cite{NK2000,KLMV}, which were based on Refs. \cite{NKGS1,NKGS2,LOS} and used fixed-order expansions. In Ref. \cite{CCMMN} the predictions with minimal prescription are small for both Tevatron and LHC energies and they account for only a very small fraction of the NNLO corrections. In contrast the predictions in Ref. \cite{PRD68,NKRV08} and the NNLL predictions in Refs. \cite{NKtt,NKtopy}, using fixed-order expansions, predicted the NNLO corrections, which are large, extremely well. 

\subsection{NLO, NNLO, and N$^3$LO expansions}

Here we present expansions of the resummed cross section through N$^3$LO. We use the $D_k$ notation for the logarithmic plus distributions of Eqs. (\ref{plus1PI}) and (\ref{plusPIM}). We give results for 1PI kinematics first. 

The NLO soft-gluon corrections from the expansion of the resummed cross section are given by 
\beq
{\hat{\sigma}}^{(1)} = \sigma^B \frac{\alpha_s(\mu_R)}{\pi}
\left\{c_3\, {\cal D}_1(s_4) + c_2\,  {\cal D}_0(s_4) \right\}
+\frac{\alpha_s^{d_{\alpha_s}+1}(\mu_R)}{\pi} A^c \, {\cal D}_0(s_4) 
\eeq
where $\sigma^B$ is the LO term. The LL coefficient is 
\beq
c_3=\sum_i 2 \, A_i^{(1)} -\sum_j A_j^{(1)} \, ,
\eeq
and it multiplies $\sigma^B$. The NLL   
terms are in general not all proportional to $\sigma^B$.
The coefficient $c_2$ is defined by $c_2=c_2^{\mu}+T_2$, 
with
$c_2^{\mu}=-\sum_i A_i^{(1)} \ln(\mu_F^2/m_t^2)$
denoting the terms involving logarithms of the scale, and  
\beqa
T_2&=&\sum_i \left[
-2 \, A_i^{(1)} \, \ln\left(\frac{-t_i}{m_t^2}\right)+D_i^{(1)}
-A_i^{(1)} \ln\left(\frac{m_t^2}{s}\right)\right]
\nonumber \\ &&
{}+\sum_j \left[B_j^{(1)}+D_j^{(1)}
-A_j^{(1)} \, \ln\left(\frac{m_t^2}{s}\right)\right] \, , 
\eeqa
where $t_i$ stands for $t-m_t^2$ or $u-m_t^2$,
while $A^c$ is defined by 
\beq
A^c={\rm tr} \left(H^{(0)} \Gamma_S^{(1)\,\dagger} S^{(0)}
+H^{(0)} S^{(0)} \Gamma_S^{(1)}\right) \, .
\eeq

The NNLO soft-gluon corrections from the expansion of the resummed cross section are 
\beqa
{\hat{\sigma}}^{(2)}&=&\sigma^B \frac{\alpha_s^2(\mu_R)}{\pi^2}
\left\{\frac{1}{2}c_3^2\, {\cal D}_3(s_4) + 
\left[\frac{3}{2}c_3 c_2-\frac{\beta_0}{4} c_3
+\sum_j \frac{\beta_0}{8} A_j^{(1)}\right]  {\cal D}_2(s_4) +\cdots \right\}
\nonumber \\ && 
{}+\frac{\alpha_s^{d_{\alpha_s}+2}(\mu_R)}{\pi^2} 
\left\{\frac{3}{2} c_3 A^c \, {\cal D}_2(s_4)+\cdots\right\}
\eeqa
where for brevity we do not show further subleading terms.

The N$^3$LO soft-gluon corrections from the expansion of the resummed cross section are 
\beqa
{\hat{\sigma}}^{(3)}&=&\sigma^B \frac{\alpha_s^3(\mu_R)}{\pi^3}
\left\{\frac{1}{8}c_3^3\, {\cal D}_5(s_4) + \left[\frac{5}{8} \, c_3^2 \, c_2 
-\frac{5}{24} \, c_3^2 \, \beta_0+\frac{5}{48} c_3 \beta_0 \sum_j A_j^{(1)} 
\right] {\cal D}_4(s_4) + \cdots \right\}
\nonumber \\ && 
{}+\frac{\alpha_s^{d_{\alpha_s}+3}(\mu_R)}{\pi^3} 
\left\{\frac{5}{8} \, c_3^2 \, A^c \; {\cal D}_4(s_4) +\cdots \right\}
\eeqa
where, again for brevity, we do not show further subleading terms.

Results for PIM kinematics are analogous: we replace $D_k(s_4)$ by $D_k(1-z)$ 
and drop terms involving the $t_i$ variables in the above expressions.

\section{Soft anomalous dimensions in top-pair production}

\subsection{Cusp anomalous dimension}

\begin{figure}
\centerline{\includegraphics[width=8cm]{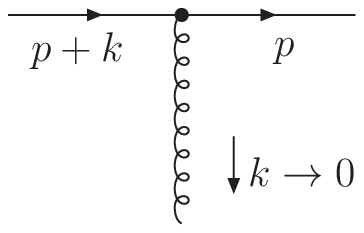}}
\caption{Eikonal diagram for soft-gluon emission from a quark.}
\label{eikdiag}
\end{figure}

The Feynman rules for diagrams with soft gluon emission 
simplify in the eikonal approximation (see Fig. \ref{eikdiag}), as
\beqa
&& {\bar u}(p) \, (-i g_s T_F^c) \, \gamma^{\mu}
\frac{i (p\!\!/+k\!\!/+m)}{(p+k)^2
-m^2+i\epsilon} 
\nonumber \\ &&
\rightarrow {\bar u}(p)\,  g_s T_F^c \, \gamma^{\mu}
\frac{p\!\!/+m}{2p\cdot k+i\epsilon}
={\bar u}(p)\, g_s T_F^c \,
\frac{v^{\mu}}{v\cdot k+i\epsilon}
\eeqa
where ${\bar u}$ is a Dirac spinor, $T_F^c$ are the generators of SU(3), and 
$p^{\mu}=(\sqrt{s}/2) v^{\mu}$.

We begin with the massive cusp anomalous dimension \cite{NK2loop}, which is 
the simplest soft anomalous dimension and an essential 
component of calculations for top-quark production.
Calculations can be performed in momentum space and in Feynman gauge, although 
other choices are possible. 

\begin{figure}
\centerline{\includegraphics[width=8cm]{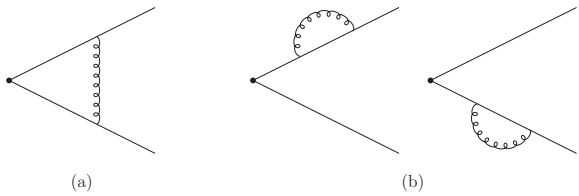}}
\caption{One-loop vertex (left) and self-energy (right) eikonal diagrams for the cusp anomalous dimension.}
\label{loopg1}
\end{figure}

The one-loop diagrams for the cusp anomalous dimension, with eikonal lines 
representing the top and the antitop quarks,  are shown in Fig. \ref{loopg1}.  
The left graph (a) is the one-loop vertex correction, while the graph on the 
right (b) shows the one-loop top and antitop self-energy diagrams.

The one-loop cusp anomalous dimension, $\Gamma_{\rm cusp}^{(1)}$, is found 
from the coefficient of the ultraviolet pole of the one-loop diagrams 
\cite{NK2loop}: 
\beq
\Gamma_{\rm cusp}^{(1)}=C_F \left[-\frac{(1+\beta^2)}{2\beta} 
\ln\left(\frac{1-\beta}{1+\beta}\right) -1\right]
\eeq
where, as before, $\beta=\sqrt{1-4m_t^2/s}$.

In terms of the cusp angle $\theta$ \cite{KorRad}, where   
$\theta=\cosh^{-1}(v_i\cdot v_j/\sqrt{v_i^2 v_j^2})=\ln[(1+\beta)/(1-\beta)]$,  
or $\beta=\tanh(\theta/2)$,
we can rewrite the one-loop expression as 
\beq
\Gamma_{\rm cusp}^{(1)}=C_F (\theta \coth\theta-1) \, .  
\eeq

\begin{figure}
\centerline{\includegraphics[width=8cm]{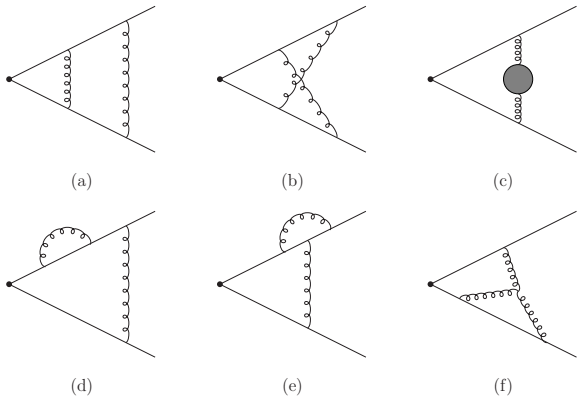}}
\caption{Two-loop vertex diagrams for the cusp anomalous dimension.}
\label{loopg2}
\end{figure}

\begin{figure}
\centerline{\includegraphics[width=8cm]{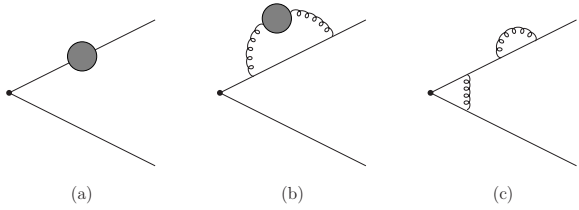}}
\caption{Two-loop top-quark self-energy graphs.}
\label{loopg2s}
\end{figure}

At two loops, the vertex-correction graphs for the cusp anomalous dimension are shown in Fig. \ref{loopg2} while the top-quark self-energy graphs are shown in Fig. \ref{loopg2s}. The grey blobs in the diagrams represent quark, gluon, and ghost loops.

With the inclusion of counterterms and after multiplying with the relevant
color factors, the two-loop cusp anomalous dimension, as determined from the 
ultraviolet poles, is \cite{NK2loop}: 
\beqa
\Gamma_{\rm cusp}^{(2)}&=&\frac{K}{2} \, \Gamma_{\rm cusp}^{(1)}
+C_F C_A \left\{\frac{1}{2}+\frac{\zeta_2}{2}+\frac{\theta^2}{2} \right.
\nonumber \\ && \hspace{30mm} 
{}-\frac{1}{2}\coth^2\theta\left[\zeta_3-\zeta_2\theta-\frac{\theta^3}{3}
-\theta \, {\rm Li}_2\left(e^{-2\theta}\right)
-{\rm Li}_3\left(e^{-2\theta}\right)\right] 
\nonumber \\ && \hspace{8mm} \left.
{}-\frac{1}{2} \coth\theta\left[\zeta_2+\zeta_2\theta+\theta^2
+\frac{\theta^3}{3}+2\, \theta \, \ln\left(1-e^{-2\theta}\right)
-{\rm Li}_2\left(e^{-2\theta}\right)\right] \right\}
\eeqa
where $\zeta_3=1.2020569\cdots$.

An excellent approximation to the complete two-loop result that is valid for all $\beta$ is given by \cite{NK2loop}:
\beq
\Gamma^{(2)}_{{\rm cusp} \, {\rm approx}}=
\frac{K}{2} \Gamma_{\rm cusp}^{(1)}+C_F C_A \left(1-\frac{2}{3}\zeta_2\right) \beta^2 \, .
\eeq

The complete three-loop result $\Gamma^{(3)}_{\rm cusp}$ is very long \cite{GHKM,NK3loop}. A very simple but excellent numerical approximation, for $n_f=5$, that is valid for all $\beta$ is given by \cite{NK3loop}
\beq
\Gamma^{(3)}_{{\rm cusp} \, {\rm approx}}=2.80322 \, \Gamma_{\rm cusp}^{(1)}+0.09221 \, \beta^2\, .
\eeq

\subsection{Soft anomalous dimension matrices for $t{\bar t}$ production}

\begin{figure}
\centerline{\includegraphics[width=2.5cm]{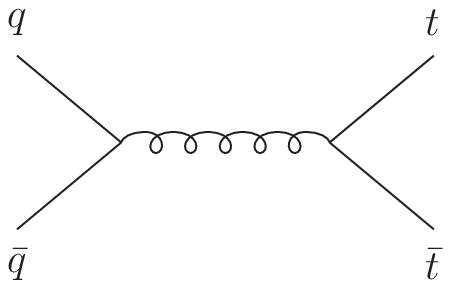}
\hspace{10mm}\includegraphics[width=9cm]{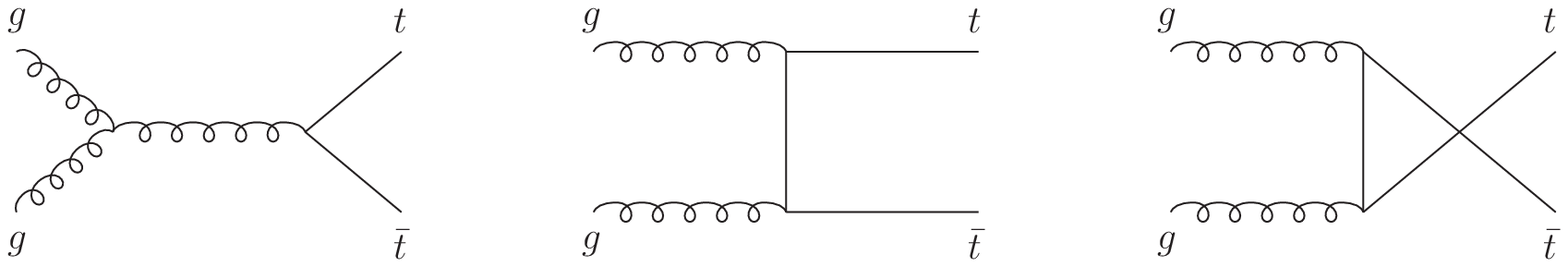}}
\caption{Lowest-order diagrams for the $q{\bar q} \rightarrow t{\bar t}$ 
channel (left diagram) and the $gg \rightarrow t{\bar t}$ channel 
(right three diagrams).}
\label{qqggdiag}
\end{figure}

The top-antitop pair production partonic processes at lowest order are
\beq
q(p_1)+{\bar q}(p_2) \rightarrow t(p_3)+ {\bar t}(p_4)
\eeq
and
\beq
g(p_1)+g(p_2) \rightarrow t(p_3)+{\bar t}(p_4) \, .
\eeq

The diagrams for these processes are shown in Fig. \ref{qqggdiag}.
We define $s=(p_1+p_2)^2$,  $t_1=(p_1-p_3)^2-m_t^2$, and $u_1=(p_2-p_3)^2-m_t^2$. 

Next, we  present the one-loop  and two-loop results for the soft anomalous matrices for these partonic processes. 
The soft anomalous dimension matrix for $q(p_1)+{\bar q}(p_2) \rightarrow t(p_3)+{\bar t}(p_4)$ in a color tensor basis of $s$-channel singlet and octet exchange,
\beq
c_1 = \delta_{12}\delta_{34}\, , \quad \quad
c_2 =  T^c_{F\; 21} \, T^c_{F\; 34}\, ,
\eeq
can be written as
\beqa
\Gamma_S^{ q{\bar q}\rightarrow t{\bar t}}=\left[\begin{array}{cc}
\Gamma^{q{\bar q}}_{11} & \Gamma^{q{\bar q}}_{12} \\
\Gamma^{q{\bar q}}_{21} & \Gamma^{q{\bar q}}_{22}
\end{array}
\right] \, .
\eeqa

At one loop we have \cite{NKGS1,NKGS2,NKhq13}
\beqa
\Gamma^{q{\bar q}\, (1)}_{11}&=&\Gamma_{\rm cusp}^{(1)} 
\nonumber \\
\Gamma^{q{\bar q}\, (1)}_{12}&=&
\frac{C_F}{C_A} \ln\left(\frac{t_1}{u_1}\right) 
\nonumber \\
\Gamma^{q{\bar q}\, (1)}_{21}&=&
2\ln\left(\frac{t_1}{u_1}\right) 
\nonumber \\ 
\Gamma^{q{\bar q}\, (1)}_{22}&=&\left(1-\frac{C_A}{2C_F}\right)
\Gamma_{\rm cusp}^{(1)} 
+4C_F \ln\left(\frac{t_1}{u_1}\right)
-\frac{C_A}{2}\left[1+\ln\left(\frac{s m_t^2 t_1^2}{u_1^4}\right)\right] \, .
\eeqa

At two loops we have \cite{NKtt,NKhq13}
\beqa
\Gamma^{q{\bar q}\,(2)}_{11}&=&\Gamma_{\rm cusp}^{(2)}
\nonumber \\ 
\Gamma^{q{\bar q}\,(2)}_{12}&=&
\left(\frac{K}{2}-\frac{C_A}{2} N_{2l}\right) \Gamma^{q{\bar q} \,(1)}_{12}
\nonumber \\
\Gamma^{q{\bar q} \,(2)}_{21}&=&
\left(\frac{K}{2}+\frac{C_A}{2} N_{2l}\right) \Gamma^{q{\bar q} \,(1)}_{21}
\nonumber \\
\Gamma^{q{\bar q} \,(2)}_{22}&=&
\frac{K}{2} \Gamma^{q{\bar q} \,(1)}_{22}
+\left(1-\frac{C_A}{2C_F}\right)
\left(\Gamma_{\rm cusp}^{(2)}-\frac{K}{2}\Gamma_{\rm cusp}^{(1)}\right) \, .
\eeqa
Here $N_{2l}$ is given by 
\beq
N_{2l}=\frac{\theta^2}{2}-\frac{1}{2} \coth\theta 
\left[\theta^2+2\, \theta \, \ln\left(1-e^{-2\theta}\right)
-{\rm Li}_2\left(e^{-2\theta}\right)\right] \, .
\eeq

The soft anomalous dimension matrix for $g(p_1)+g(p_2) \rightarrow t(p_3)+{\bar t}(p_4)$ in a color tensor basis
\beq
c_1=\delta^{12}\,\delta_{34}, \quad c_2=d^{12c}\,T^c_{34},
\quad c_3=i f^{12c}\,T^c_{34} 
\eeq
where $d$ and $f$ are the totally symmetric and 
antisymmetric $SU(3)$ tensors, is
\beqa
\Gamma_S^{gg\rightarrow t{\bar t}}=\left[\begin{array}{ccc}
\Gamma^{gg}_{11} & 0 & \Gamma^{gg}_{13} \vspace{2mm} \\
0 & \Gamma^{gg}_{22} & \Gamma^{gg}_{23} \vspace{2mm} \\
\Gamma^{gg}_{31} & \Gamma^{gg}_{32} & \Gamma^{gg}_{22}
\end{array}
\right] \, .
\eeqa

At one loop we have \cite{NKGS2,NKhq13}
\beqa
\Gamma^{gg\,(1)}_{11}&=& \Gamma_{\rm cusp}^{(1)}
\nonumber \\
\Gamma^{gg\,(1)}_{13}&=& \ln\left(\frac{t_1}{u_1}\right) 
\nonumber \\
\Gamma^{gg\,(1)}_{31}&=& 2 \ln\left(\frac{t_1}{u_1}\right) 
\nonumber \\
\Gamma^{gg\,(1)}_{22}&=& \left(1-\frac{C_A}{2C_F}\right)
\Gamma_{\rm cusp}^{(1)}
-\frac{C_A}{2}\left[1+\ln\left(\frac{s m_t^2}{t_1 u_1}\right)\right] 
\nonumber \\
\Gamma^{gg\,(1)}_{23}&=&\frac{C_A}{2} \ln\left(\frac{t_1}{u_1}\right) 
\nonumber \\
\Gamma^{gg\,(1)}_{32}&=&\frac{(N_c^2-4)}{2N_c} \ln\left(\frac{t_1}{u_1}\right) 
\eeqa

At two loops we find \cite{NKtt,NKhq13}
\beqa
\Gamma^{gg\,(2)}_{11}&=& \Gamma_{\rm cusp}^{(2)}
\nonumber \\
\Gamma^{gg\,(2)}_{13}&=&\left(\frac{K}{2}-\frac{C_A}{2} N_{2l}\right) 
\Gamma^{gg \,(1)}_{13} 
\nonumber \\
\Gamma^{gg\,(2)}_{31}&=&\left(\frac{K}{2}+\frac{C_A}{2} N_{2l}\right)  
\Gamma^{gg \,(1)}_{31} 
\nonumber \\
\Gamma^{gg\,(2)}_{22}&=& \frac{K}{2} \Gamma^{gg \,(1)}_{22}
+\left(1-\frac{C_A}{2C_F}\right)
\left(\Gamma_{\rm cusp}^{(2)}-\frac{K}{2}\Gamma_{\rm cusp}^{(1)}\right)
\nonumber \\
\Gamma^{gg\,(2)}_{23}&=&\frac{K}{2} \Gamma^{gg \,(1)}_{23}
\nonumber \\
\Gamma^{gg\,(2)}_{32}&=&\frac{K}{2} \Gamma^{gg \,(1)}_{32}  
\eeqa

\section{Top-antitop pair production}

\subsection{Total cross sections for $t{\bar t}$ production}

\begin{table}[htb]
\begin{center}
\begin{tabular}{|c|c|} \hline
Collider Energy & Cross section $\pm$ scale $\pm$ pdf  \\ \hline
1.8 TeV $p{\bar p}$ &  $6.130^{+0.076}_{-0.233}{}^{+0.179}_{-0.160}$ 
\\ \hline
1.96 TeV $p{\bar p}$ & $7.876^{+0.096}_{-0.293}{}^{+0.224}_{-0.203}$
\\ \hline
5.02 TeV $pp$ &  $71.3^{+2.2}_{-3.3}{}^{+1.9}_{-2.8}$
\\ \hline
7 TeV $pp$  &  $183.0^{+5.4}_{-6.9}{}^{+4.0}_{-5.6}$
\\ \hline
 8 TeV $pp$ &  $260.1^{+7.4}_{-8.6}{}^{+5.3}_{-7.3}$
\\ \hline
13 TeV $pp$ &  $842.5^{+25.3}_{-16.9}{}^{+13.7}_{-17.7}$
\\ \hline
14 TeV $pp$ &  $995.0^{+29.7}_{-19.0}{}^{+15.7}_{-20.2}$
\\ \hline
\end{tabular}
\caption[]{aN$^3$LO top-antitop production cross sections \cite{N3LO1} with $m_t=172.5$ GeV at LHC $pp$ and Tevatron $p{\bar p}$ collider energies.}
\label{topantitop}
\end{center}
\end{table}

We begin our presentation of numerical results with the total cross section 
for top-antitop pair production. The total hadronic cross section is calculated by integrating over the convolution of the double-differential partonic cross section with the parton distribution functions $\phi$.

We denote the NLO soft-gluon corrections from the expansion of the NNLL resummed
cross section as approximate NLO (aNLO) corrections. Similarly the NNLO soft-gluon corrections are denoted as approximate NNLO (aNNLO) corrections, and the 
N$^3$LO soft-gluon corrections are denoted as approximate N$^3$LO (aN$^3$LO) corrections. The aNLO and aNNLO corrections are extremely good approximations to the exact NLO and NNLO results, respectively, for total cross sections as well as top-quark differential distributions at all Tevatron and LHC energies (see e.g. the discussion in Ref. \citen{NKhq13}). 
The best aN$^3$LO prediction is given by the sum of the NNLO cross section and the aN$^3$LO soft-gluon corrections.

\begin{figure}
\centerline{\includegraphics[width=13cm]{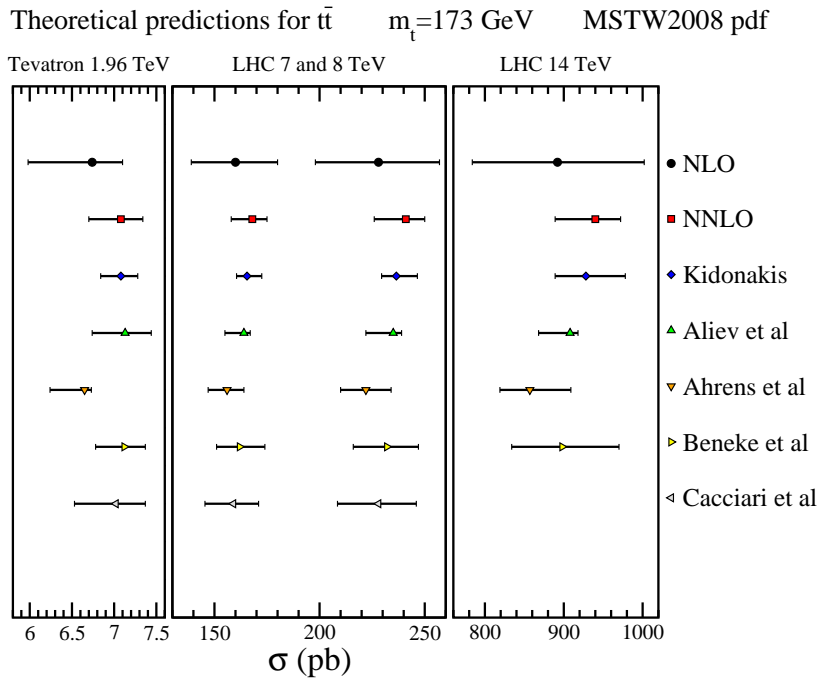}}
\caption{Theoretical predictions with soft-gluon corrections \cite{NKtt,ALLMUW,AFNPY4,BFKS,CCMMN} from 2010 and 2011 compared to NLO and NNLO results for Tevatron and LHC energies.}
\label{ttbarpredictions}
\end{figure}

A comparison of various approximate predictions using higher-order soft-gluon corrections, all made before the exact NNLO cross section was known, is shown in Fig. \ref{ttbarpredictions}, all with the same choice of parameters and MSTW2008 \cite{MSTW2008} pdf, at 1.96 TeV Tevatron energy and at 7, 8, and 14 TeV LHC energies. Moreover, exact NLO and NNLO results for the total cross sections are also shown on the plot. We observe the success or lack thereof of the various predictions in predicting the exact NNLO result.

Ref. \cite{NKtt} uses the QCD moment-space resummation formalism for the double-differential cross section. Ref. \cite{ALLMUW} uses QCD moment-space resummation with absolute threshold for the total-only cross section. Ref. \cite{AFNPY1} uses the SCET resummation formalism for the double-differential cross section, while Ref. \cite{BFKS} uses SCET resummation with absolute threshold for the total-only cross section. Lastly, Ref. \cite{CCMMN}  uses QCD moment-space resummation with absolute threshold for the total-only cross section. 

The result in Ref. \cite{NKtt} is very close to the exact NNLO \cite{NNLOtt1} result: the central values and the scale uncertainty are nearly identical, for all collider energies, with less than 1\% difference between approximate and exact cross sections.
This was expected from the comparison of approximate NNLO results in different kinematics in Ref. \cite{PRD68} (see also the discussions in \cite{NKtt} and \cite{NKhq13}).

\begin{figure}
\centerline{\includegraphics[width=12cm]{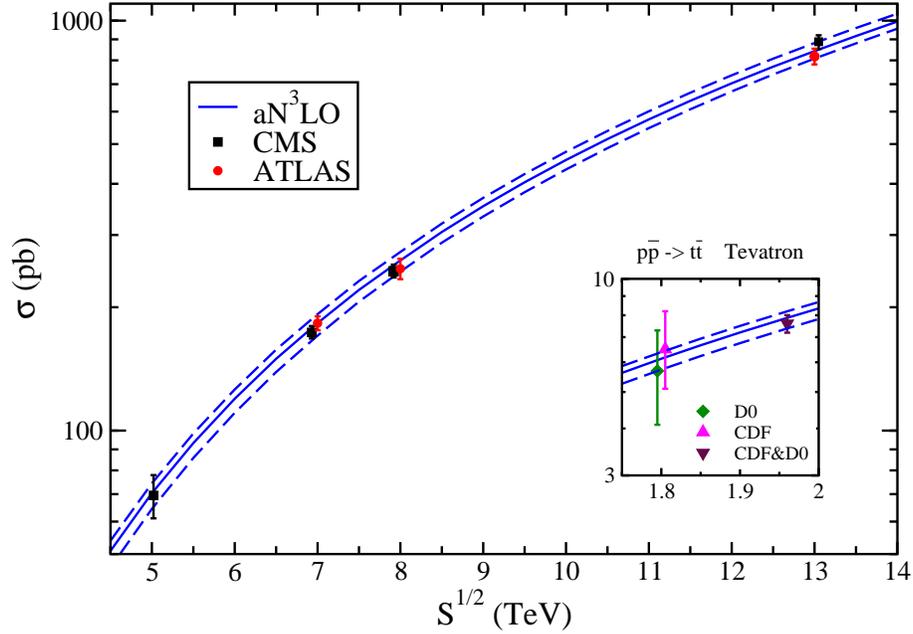}}
\caption{The aN$^3$LO \cite{N3LO1} top-antitop pair total cross section, with theoretical uncertainties, as a function of LHC energy, compared with CMS data at 5.02 TeV \cite{CMS5.02} and with ATLAS and CMS data at 7 TeV \cite{ATLAStt7,CMStt7and8}, 8 TeV \cite{ATLAStt8,CMStt7and8}, and 13 TeV \cite{ATLAStt13,CMStt13} LHC energies. The inset plot shows the aN$^3$LO cross section at Tevatron energies compared with CDF \cite{CDF1.8} and D0 \cite{D01.8} data at 1.8 TeV, and CDF\&D0 combination \cite{CDFD01.96} at 1.96 TeV.}
\label{ttSaN3LO}
\end{figure}

In Fig. \ref{ttSaN3LO} we display theoretical predictions at aN$^3$LO\cite{N3LO1} for the total cross section as a function of top-quark mass at the LHC and the Tevatron.
We use MMHT2014 \cite{MMHT2014} NNLO pdf but note that the results with CT14\cite{CT14} and NNPDF \cite{NNPDF} pdf are very similar.
We compare the aN$^3$LO results with data from the LHC at 5.02 TeV \cite{CMS5.02}, 7 TeV \cite{ATLAStt7,CMStt7and8}, 8 TeV \cite{ATLAStt8,CMStt7and8}, and 13 TeV \cite{ATLAStt13,CMStt13}, and from the Tevatron at 1.8 TeV \cite{CDF1.8,D01.8} and 1.96 TeV \cite{CDFD01.96}. We find superb agreement between the theoretical predictions and the data. 

In Table 1 we present the aN$^3$LO cross sections\cite{N3LO1} for $t{\bar t}$ production.
The central result at each energy is with $\mu_F=\mu_R=m_t$, the first uncertainty is from independent variation of $\mu_F$ and $\mu_R$ over the range $m_t/2$ to $2m_t$, and the second uncertainty is from the MMHT2014 \cite{MMHT2014} NNLO pdf.

\subsection{Top-quark $p_T$ and rapidity distributions in $t{\bar t}$ production}

We continue with top quark differential distributions in $t{\bar t}$ 
production.
We present theoretical results for the top-quark transverse momentum and 
rapidity distributions at Tevatron and LHC energies.

\begin{figure}
\centerline{\includegraphics[width=10cm]{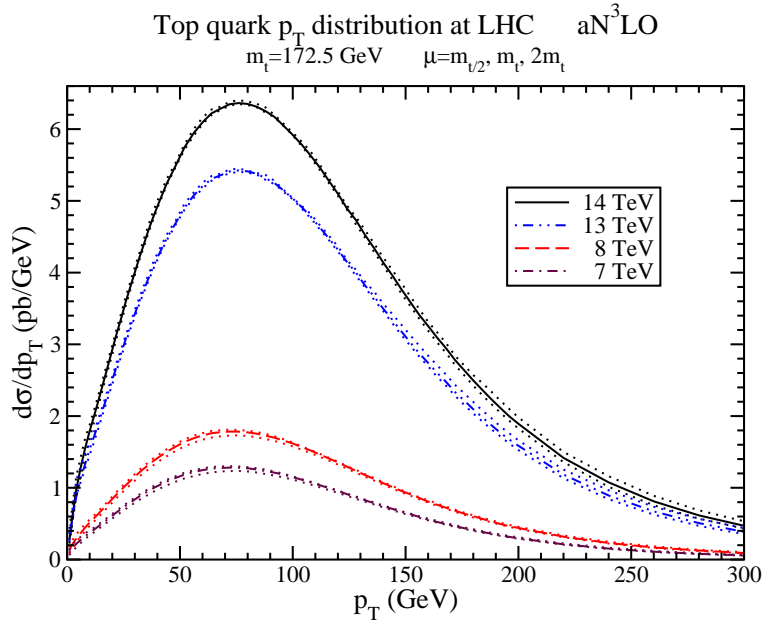}}
\caption{The aN$^3$LO \cite{N3LO2} top-quark $p_T$ distributions at 7, 8, 13, and 14 TeV LHC energies.}
\label{ptaN3LO}
\end{figure}

The aN$^3$LO top-quark $p_T$ distributions \cite{N3LO2}, with scales $\mu=m_t/2$, $m_t$, and $2m_t$, and $m_t=172.5$ GeV, are displayed in Fig. \ref{ptaN3LO} at 7, 8, 13, and 14 TeV LHC energies.

\begin{figure}
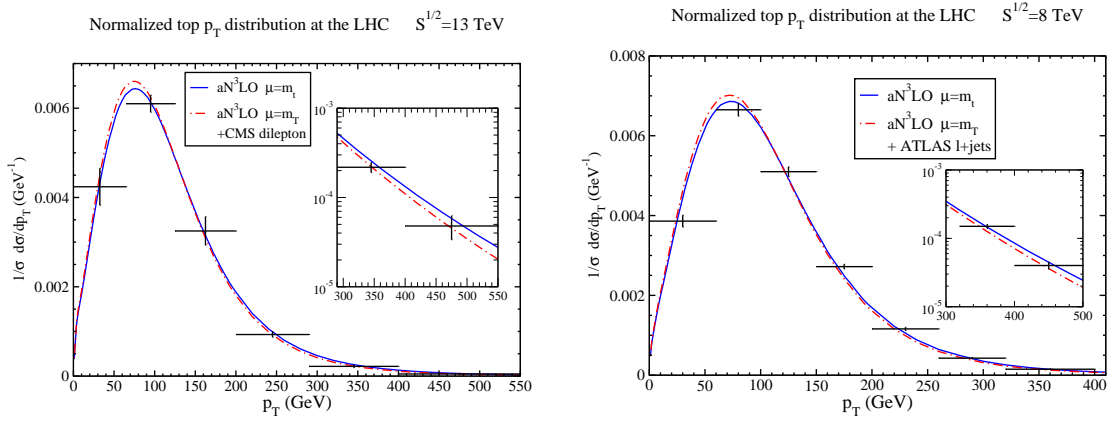

\centerline{\includegraphics[width=7cm]{ptnormaN3LO13lhcCMS1708.07638dileptplot.eps}
\hspace{5mm}\includegraphics[width=7cm]{ptnormaN3LO8lhcATLAS1511.04716leptjetplot.eps}}
\caption{The aN$^3$LO top-quark normalized $p_T$ distribution (left) at 13 TeV LHC energy compared with CMS \cite{CMSpty13} data and (right) at 8 TeV energy compared with ATLAS \cite{ATLASpty8} lepton+jets data.}
\label{pt13-8lhc}
\end{figure}

The left plot in Fig. \ref{pt13-8lhc} shows the aN$^3$LO top-quark normalized $p_T$ distribution, $(1/\sigma) d\sigma/dp_T$, at the LHC compared to CMS \cite{CMSpty13} data at 13 TeV energy. Two different choices of scale are used, $\mu=m_t$ and $\mu=m_T$, where $m_T=\sqrt{p_T^2+m_t^2}$. There is excellent agreement with data. In the right plot of Fig. \ref{pt13-8lhc}, the aN$^3$LO top-quark normalized $p_T$ distribution at 8 TeV LHC energy with scale $m_t$ and $m_T$ is compared to ATLAS \cite{ATLASpty8} data, again with excellent agreement. The high-$p_T$ region is highlighted in the inset plots.

\begin{figure}
\centerline{\includegraphics[width=7cm]{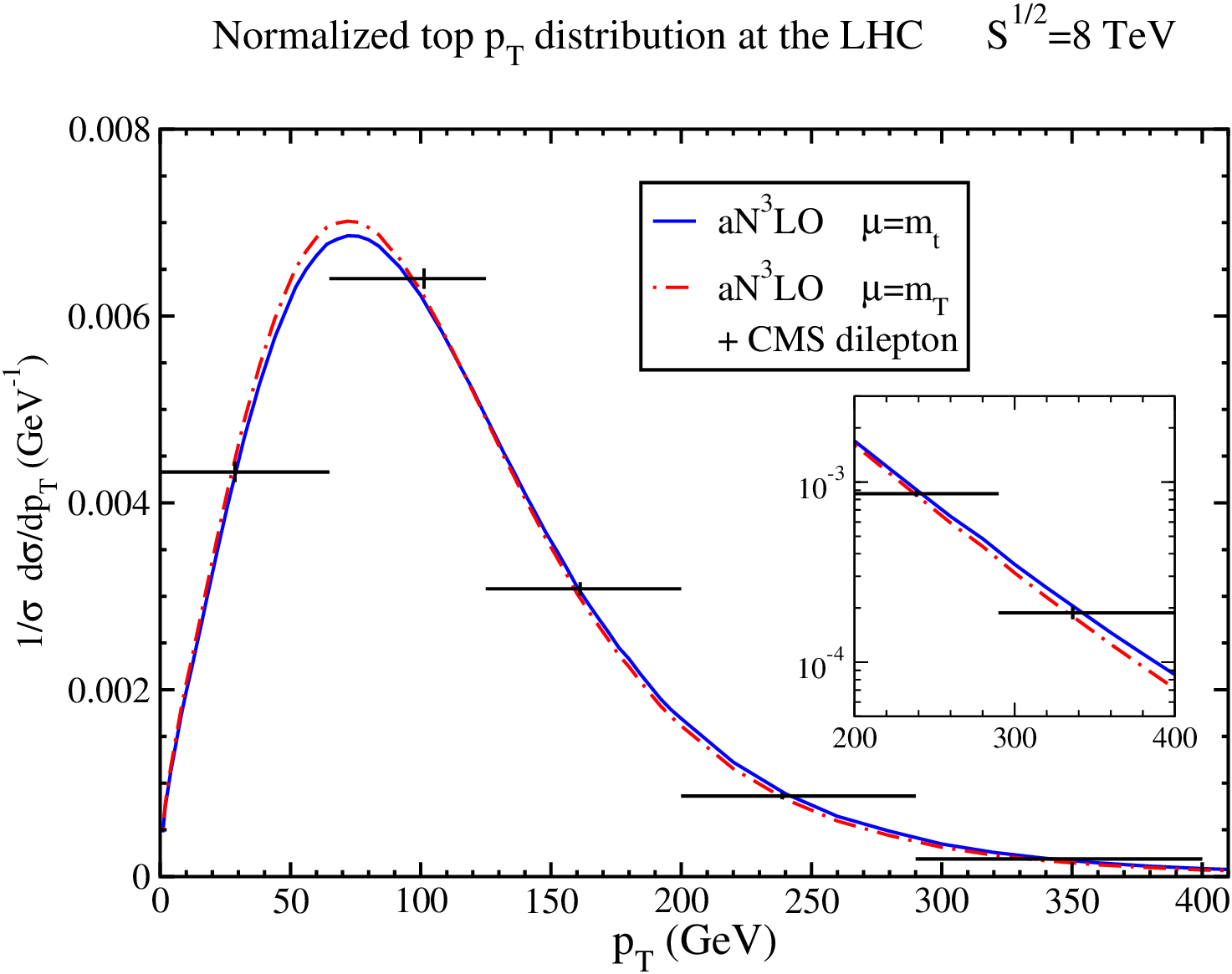}
\hspace{5mm}\includegraphics[width=7cm]{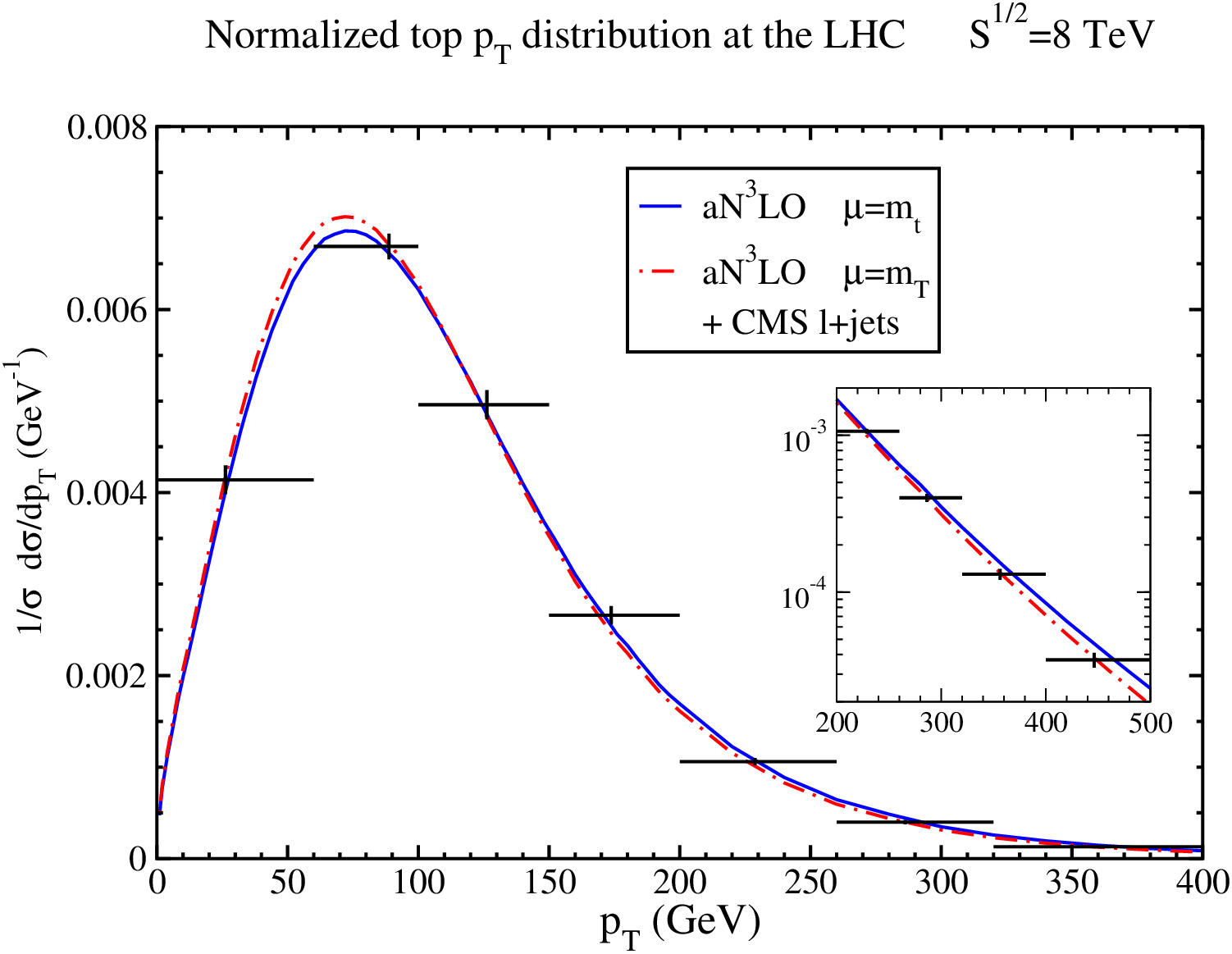}}
\caption{The aN$^3$LO top-quark normalized $p_T$ distribution at 8 TeV LHC energy compared with CMS \cite{CMSpty8} dilepton (left) and lepton+jets (right) data.}
\label{pt8lhc}
\end{figure}

Figure \ref{pt8lhc} shows the aN$^3$LO top-quark normalized $p_T$ distribution at the LHC compared to CMS \cite{CMSpty8} data at 8 TeV energy. There is excellent agreement with data in both the dilepton and lepton+jets channels.

\begin{figure}
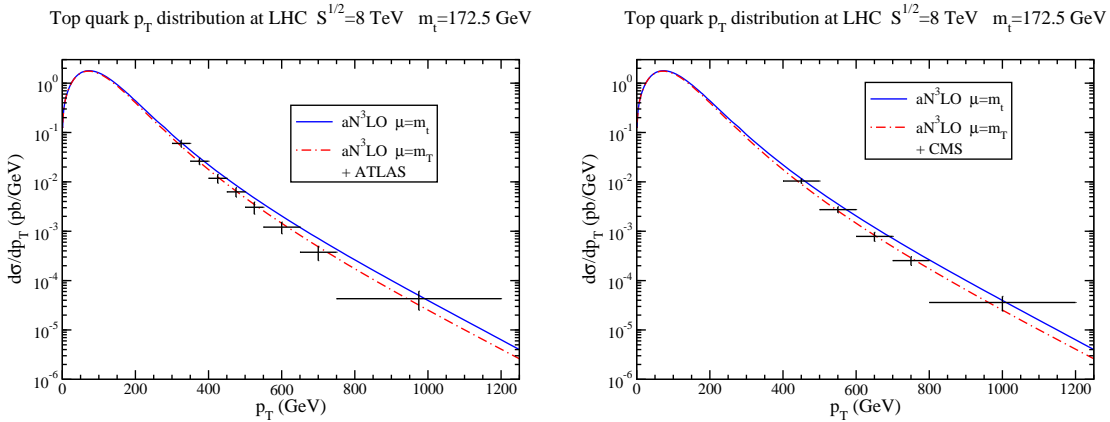

\centerline{\includegraphics[width=7cm]{ptaN3LO8lhcboostATLAS1510.03818plot.eps}
\hspace{5mm}\includegraphics[width=7cm]{ptaN3LO8lhcboostCMS1605.00116plot.eps}}
\caption{The aN$^3$LO top-quark $p_T$ distribution at 8 TeV LHC energy compared with ATLAS \cite{ATLASptboost8} (left) and CMS \cite{CMSptboost8} (right) data.}
\label{pt8lhcboost}
\end{figure}

Figure \ref{pt8lhcboost} shows the aN$^3$LO boosted-top quark $p_T$ distribution at the LHC compared to ATLAS \cite{ATLASptboost8} and CMS \cite{CMSptboost8} data at 8 TeV energy. There is excellent agreement with the data.

\begin{figure}
\centerline{\includegraphics[width=7cm]{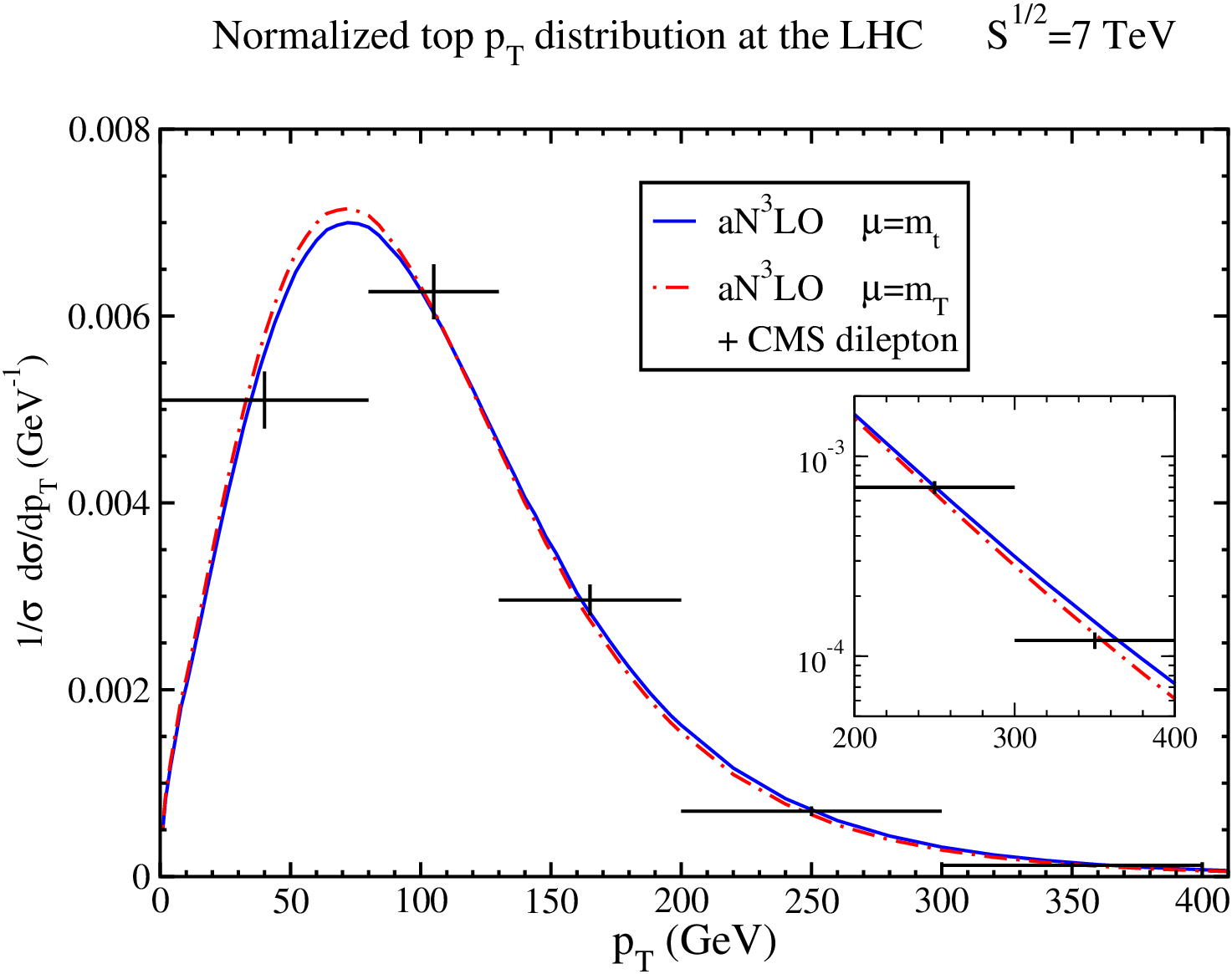}
\hspace{5mm}\includegraphics[width=7cm]{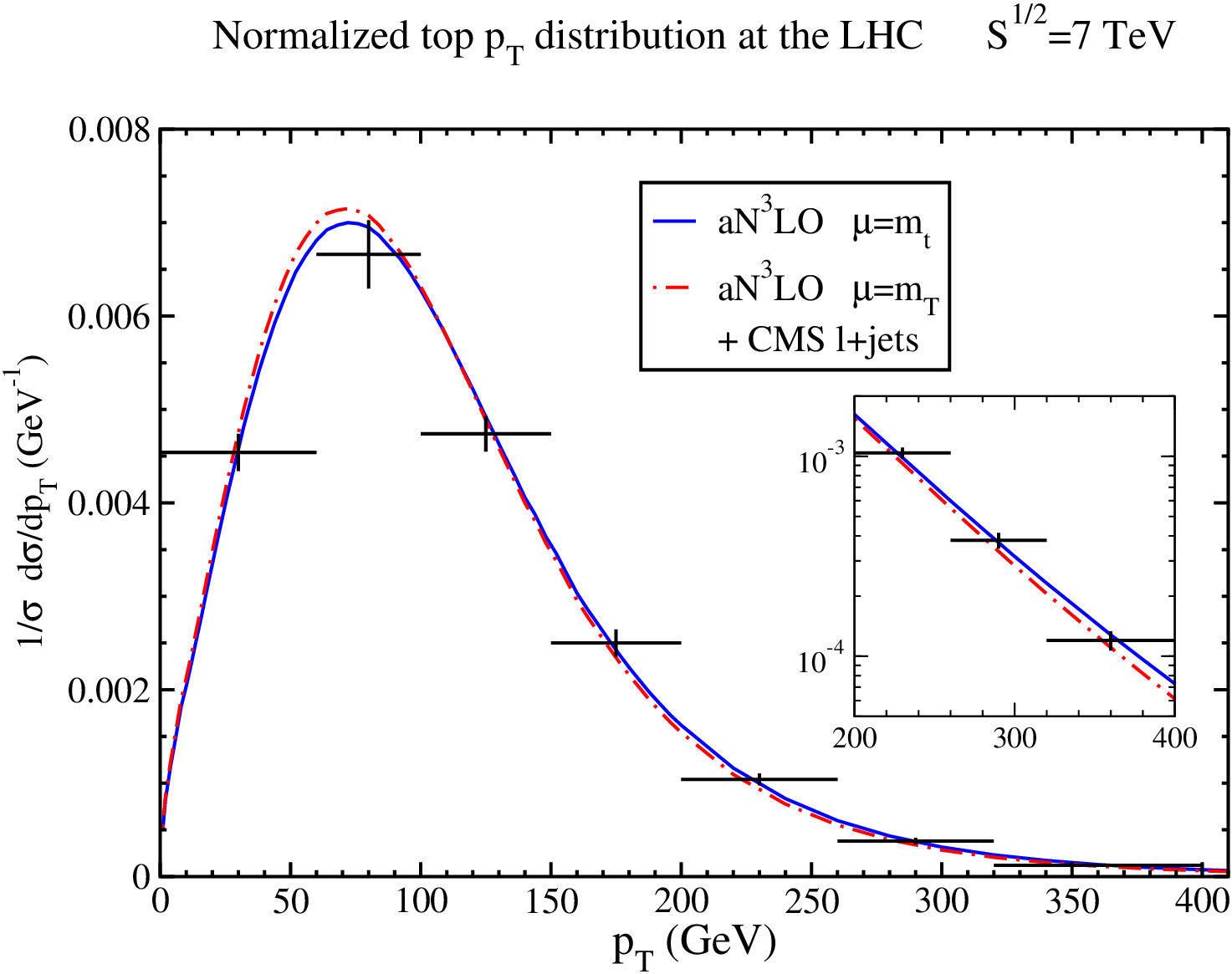}}
\caption{The aN$^3$LO top-quark normalized $p_T$ distribution at 7 TeV LHC energy compared with CMS \cite{CMSpty7} dilepton (left) and lepton+jets (right) data.}
\label{pt7lhcCMS}
\end{figure}

The aN$^3$LO top-quark normalized $p_T$ distribution at 7 TeV LHC energy is shown in Fig. \ref{pt7lhcCMS} and compared with CMS \cite{CMSpty7} data in the dilepton and lepton+jets channels. We note the excellent agreement of theory with data.

\begin{figure}
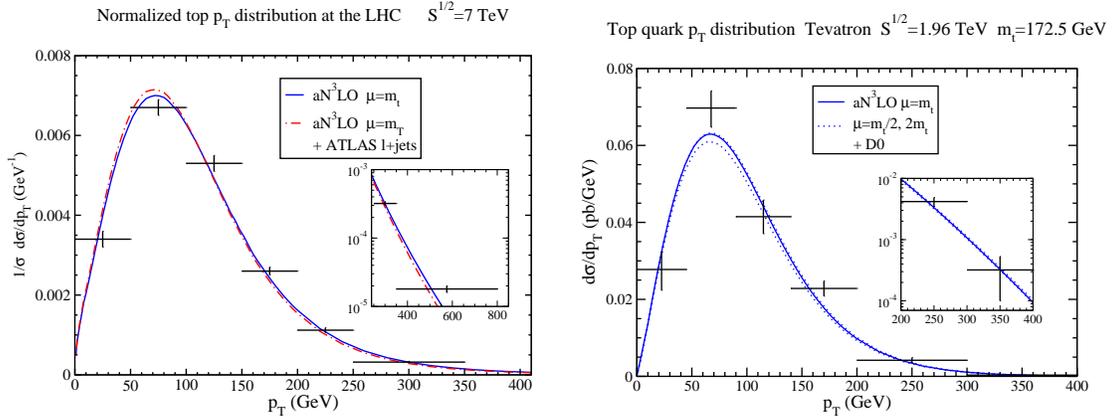

\centerline{\includegraphics[width=7cm]{ptnormaN3LO7lhcATLAS1407.0371leptjetplot.eps}
\hspace{5mm}\includegraphics[width=7cm]{ptaN3LOtevD01401.5785plot.eps}}
\caption{(Left) The aN$^3$LO top-quark normalized $p_T$ distribution at 7 TeV LHC energy compared with ATLAS \cite{ATLASpt7} lepton+jets data. (Right) The aN$^3$LO top-quark $p_T$ distribution at 1.96 TeV Tevatron energy compared with data from D0 \cite{D0pty}.}
\label{pt7lhc-tev}
\end{figure}

In the left plot of Fig. \ref{pt7lhc-tev}, the aN$^3$LO top-quark normalized $p_T$ distribution at 7 TeV LHC energy is compared to ATLAS \cite{ATLASpt7} data.
The right plot of Fig. \ref{pt7lhc-tev} displays the aN$^3$LO top quark $p_T$ distribution at 1.96 TeV Tevatron energy.
Excellent agreement of the aN$^3$LO results with D0 data \cite{D0pty} can be seen, including the high-$p_T$ region shown in the inset.

\begin{figure}
\centerline{\includegraphics[width=10cm]{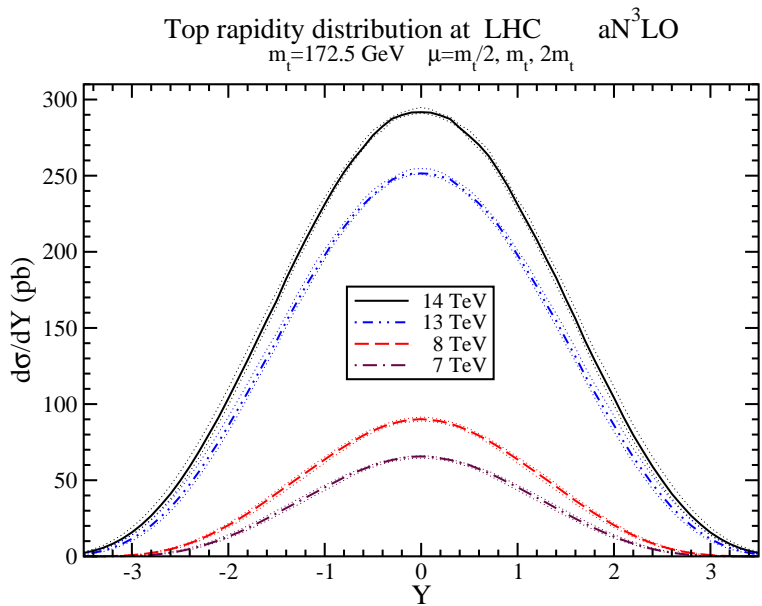}}
\caption{The aN$^3$LO \cite{N3LO2} top-quark rapidity distributions at 7, 8, 13, and 14 TeV LHC energies.}
\label{yaN3LO}
\end{figure}

The aN$^3$LO top-quark rapidity distributions at 7, 8, 13, and 14 TeV LHC energies are displayed in Fig. \ref{yaN3LO} with scale choices $\mu=m_t/2$, $m_t$, and $2m_t$.

\begin{figure}
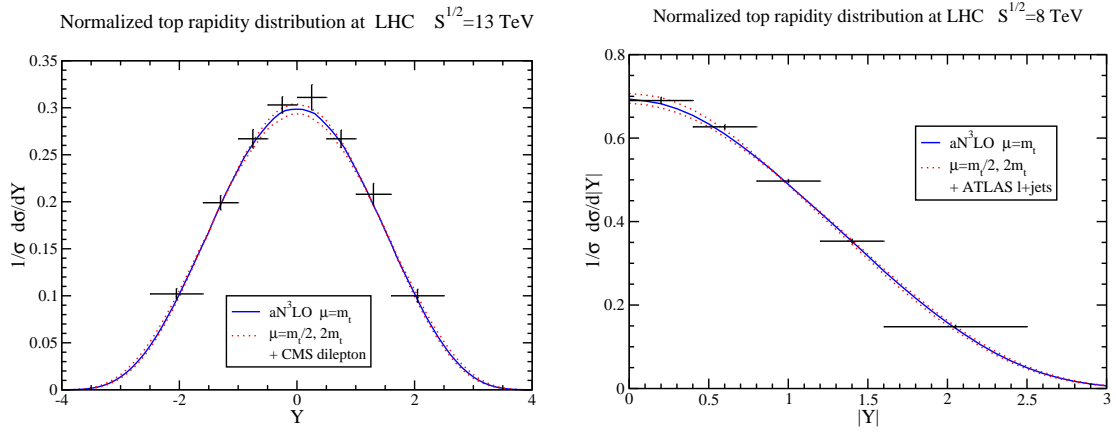

\centerline{\includegraphics[width=7cm]{ynormaN3LO13lhcCMS1708.07638dileptplot.eps}
\hspace{5mm}\includegraphics[width=7cm]{ynormabsaN3LO8lhcATLAS1511.04716leptjetplot.eps}}
\caption{(Left) The aN$^3$LO top-quark normalized rapidity distribution at 13 TeV LHC energy compared with CMS data \cite{CMSpty13}. (Right) The aN$^3$LO top-quark normalized absolute value rapidity distribution at 8 TeV LHC energy compared with ATLAS \cite{ATLASpty8} data.}
\label{y13-8lhc}
\end{figure}

The aN$^3$LO top-quark normalized rapidity distribution, $(1/\sigma) d\sigma/dY$, at 13 TeV LHC energy is shown in the left plot of Fig. \ref{y13-8lhc} and compared with CMS \cite{CMSpty13} data.
In the right plot of Fig. \ref{y13-8lhc}, the aN$^3$LO top-quark normalized absolute value rapidity distribution at 8 TeV LHC energy is compared to ATLAS \cite{ATLASpty8} data. We find excellent agreement between theory and data in both cases.

\begin{figure}
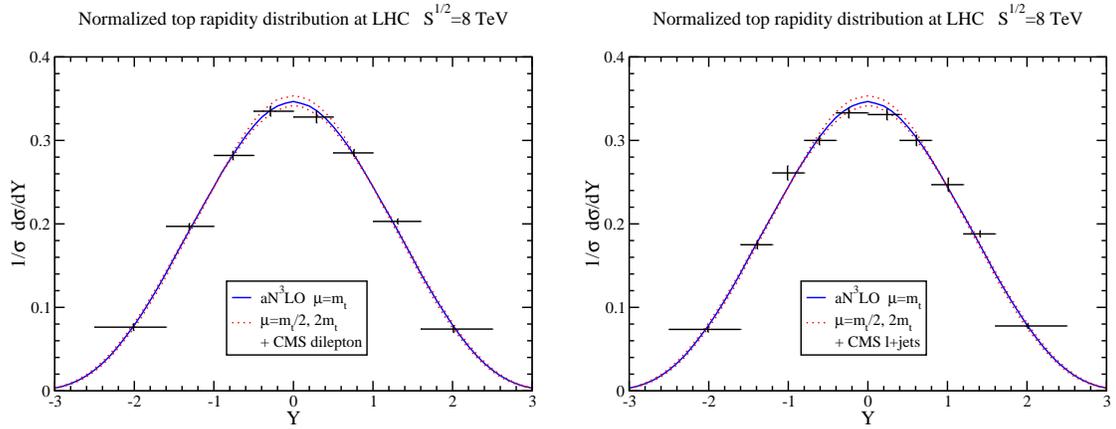

\centerline{\includegraphics[width=7cm]{ynormaN3LO8lhcCMS1505.04480dileptplot.eps}
\hspace{5mm}\includegraphics[width=7cm]{ynormaN3LO8lhcCMS1505.04480leptjetplot.eps}}
\caption{The aN$^3$LO top-quark normalized rapidity distribution at 8 TeV LHC energy compared with CMS \cite{CMSpty8} dilepton (left) and lepton+jets (right) data.}
\label{y8lhc}
\end{figure}

The aN$^3$LO top-quark normalized rapidity distribution at 8 TeV LHC energy is shown in Fig. \ref{y8lhc} and compared with CMS \cite{CMSpty8} dilepton and lepton+jets data, again with excellent agreement.

\begin{figure}
\centerline{\includegraphics[width=7cm]{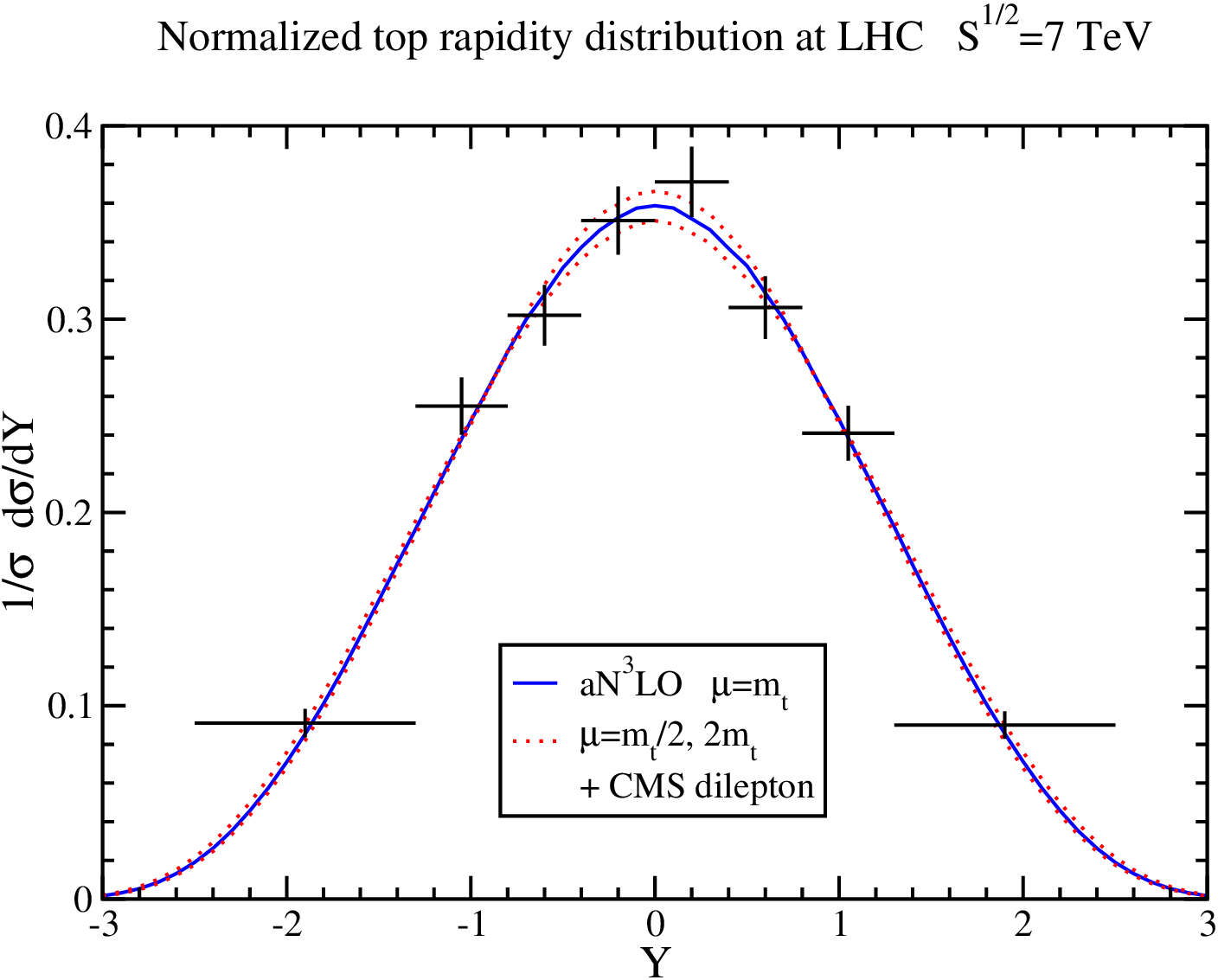}
\hspace{5mm}\includegraphics[width=7cm]{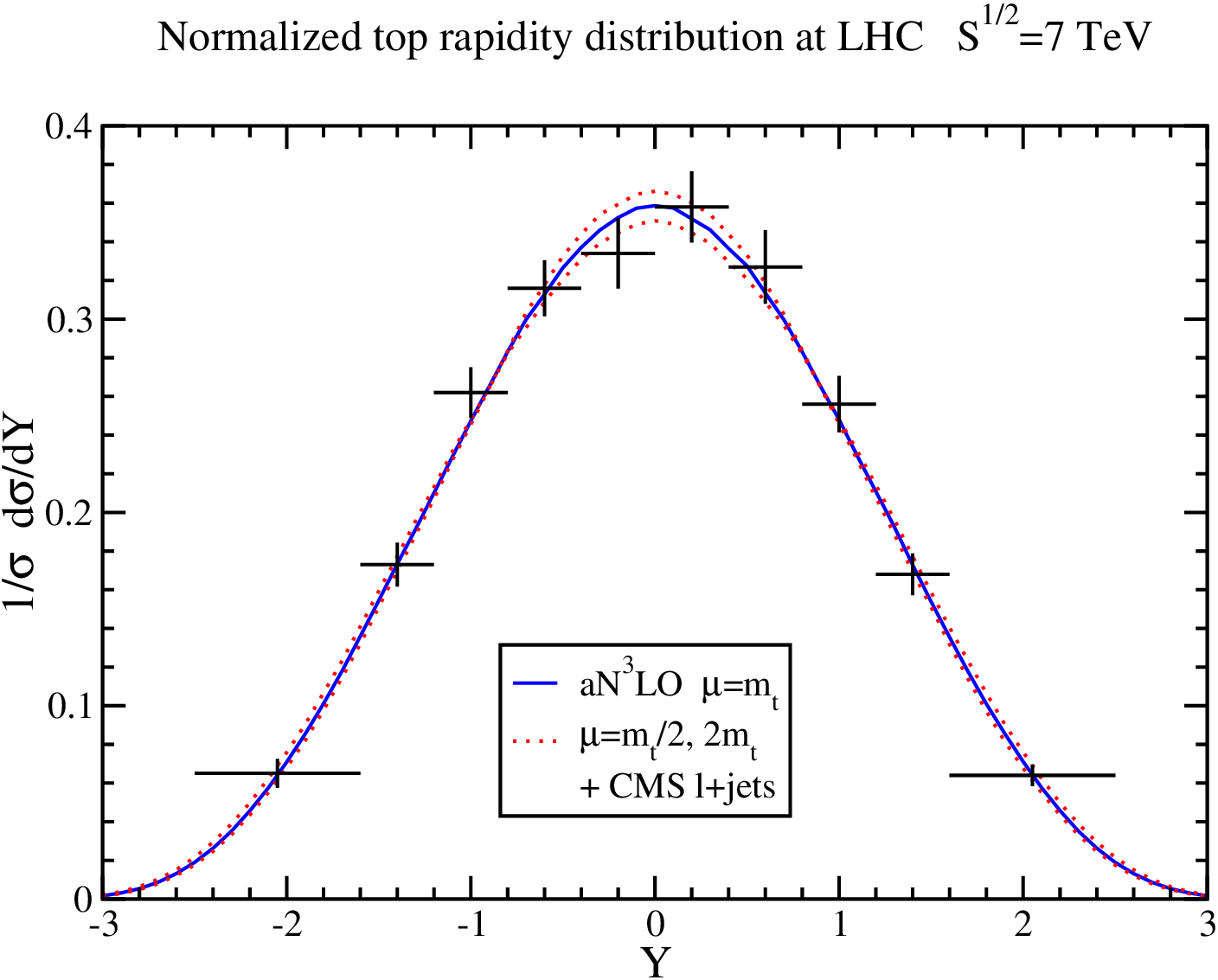}}
\caption{The aN$^3$LO top-quark normalized rapidity distribution at 7 TeV LHC energy compared with CMS \cite{CMSpty7} dilepton (left) and lepton+jets (right) data.}
\label{y7lhc}
\end{figure}

The aN$^3$LO top-quark normalized rapidity distribution at 7 TeV LHC energy is shown in Fig. \ref{y7lhc} and compared with CMS \cite{CMSpty7} dilepton and lepton+jets data. We again note the excellent agreement between theory and data.

\begin{figure}
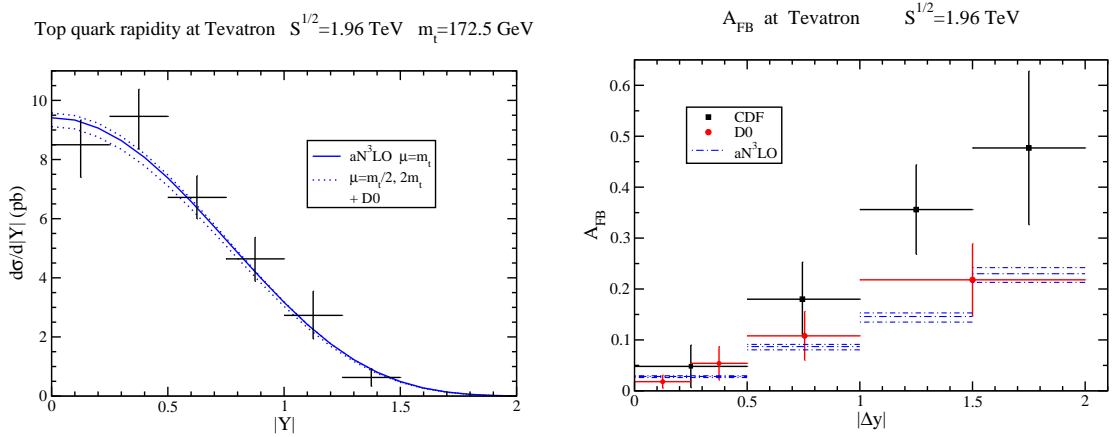

\centerline{\includegraphics[width=7cm]{yabsaN3LOtevD01401.5785plot.eps}
\hspace{5mm}\includegraphics[width=7cm]{AFBexpplot.eps}}
\caption{(Left) The aN$^3$LO top-quark rapidity distribution at 1.96 TeV Tevatron energy compared with D0 \cite{D0pty} data. (Right) The aN$^3$LO differential forward-backward asymmetry at 1.96 Tevatron energy compared with D0 \cite{D0afb} and CDF \cite{CDFafb} data.}
\label{yafbtev}
\end{figure}

The aN$^3$LO top-quark rapidity distribution has been calculated  
for 1.96 TeV Tevatron energy as shown in the left plot in 
Fig. \ref{yafbtev}, and it is in very good agreement with data from 
D0 \cite{D0pty}.

\subsection{Top-quark forward-backward asymmetry}

The top-quark forward-backward asymmetry is defined by 
\beq
A_{\rm FB}=\frac{\sigma(Y>0)-\sigma(Y<0)}{\sigma(Y>0)+\sigma(Y<0)} \, .
\eeq
The asymmetry is very significant at the Tevatron. In addition to QCD corrections, electroweak corrections are important for this asymmetry \cite{WBZS,WBS10,WHDP,AMMT} (see also Ref. \cite{PTZ} for rapidity distributions at the LHC).
The theoretical result at aN$^3$LO and including electroweak corrections for 1.96 TeV Tevatron energy is \cite{N3LO3} 
$A_{\rm FB}=0.100 \pm 0.006$ which is in agreement with the CDF and D0 
combination \cite{CDFD0afb} of $0.128 \pm 0.025$.
The aN$^3$LO differential $A_{\rm FB}$ is plotted in the right plot of Fig. \ref{yafbtev} and compared with D0 \cite{D0afb} and CDF \cite{CDFafb} data.

\section{Single-top production}

\begin{table}[htb]
\begin{center}
\begin{tabular}{|c|c|c|c|c|c|c|c|} \hline
Collider & $t$-channel & $t$-channel & $t$-channel & $s$-channel & $s$-channel & $s$-channel &$tW^-$ \\ 
Energy & $t$ &  ${\bar t}$ & $t$ and ${\bar t}$  & $t$ &  ${\bar t}$ & $t$ and ${\bar t}$ &  \\ \hline
1.96 TeV $p{\bar p}$ & 1.088  & 1.088 & 2.176 & 0.52 &  0.52  & 1.04 & 0.102
\\ \hline
7 TeV $pp$  &  43.9 & 23.7 &  67.6 & 3.21 & 1.56 &  4.77 &  8.5
\\ \hline
 8 TeV $pp$ &  57.5 & 31.7 &  89.2 & 3.86 & 1.96 &  5.82 & 12.0
\\ \hline
13 TeV $pp$ & 139.6 & 84.0 & 223.6 & 7.29 & 4.20 & 11.49 & 38.1 
\\ \hline 
14 TeV $pp$ & 158.3 & 96.1 & 254.3 & 7.98 & 4.68 & 12.66 & 44.8
\\ \hline
\end{tabular}
\caption[]{aNNLO $t$-channel\cite{NKtch} and $s$-channel \cite{NKsch} single-top, single-antitop, and combined cross sections, and aN$^3$LO $tW^-$ \cite{NKtW16} cross sections with $m_t=172.5$ GeV at LHC $pp$ and Tevatron $p{\bar p}$ collider energies.}
\label{singletop}
\end{center}
\end{table}

\begin{figure}
\centerline{\includegraphics[width=2.5cm]{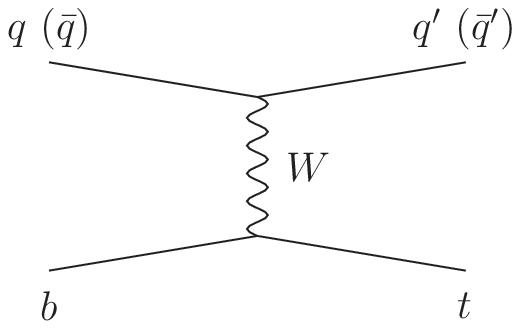}
\hspace{7mm} \includegraphics[width=2.5cm]{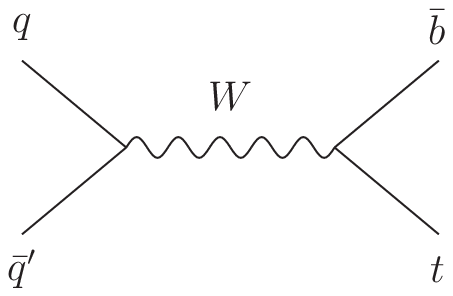}
\hspace{7mm} \includegraphics[width=6cm]{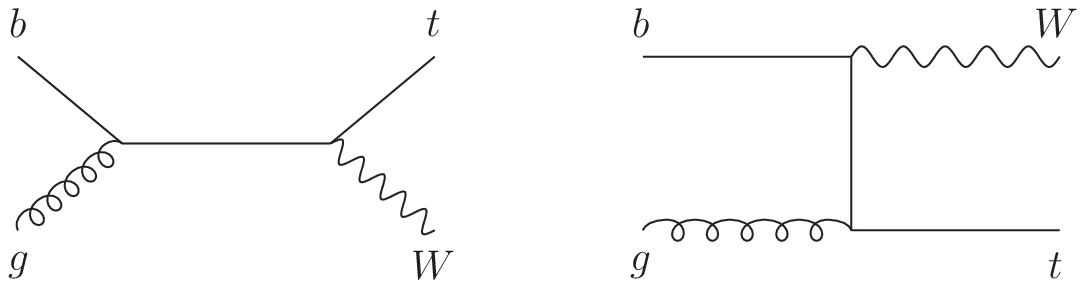}}
\caption{Lowest-order diagrams for single-top production in the $t$-channel (left diagram), $s$-channel (second from left), and in $tW$ production (right two diagrams).}
\label{singletopdiag}
\end{figure}

Single-top-quark production was first observed at the Tevatron in 2009 \cite{D0singletop,CDFsingletop}. The single-top partonic processes at lowest order are shown in Fig. \ref{singletopdiag}.

The $t$-channel partonic processes are of the form  $qb \rightarrow q' t$ and ${\bar q} b \rightarrow {\bar q}' t$ and are numerically the largest among single-top processes at Tevatron and LHC energies.
The $s$-channel partonic processes are of the form $q{\bar q}' \rightarrow {\bar b} t$ and are numerically second largest at the Tevatron and the smallest at the LHC among single-top processes.
The associated $tW$ production proceeds via $bg \rightarrow tW^-$
and is negligible at the Tevatron but second largest numerically among single-top processes at the LHC.

Table 2 summarizes the central values of the cross sections for the various single-top channels at LHC and Tevatron energies.

\subsection{$t$-channel production}

We start with single-top production in the $t$-channel. 

The soft anomalous dimension matrix for $t$-channel single-top production is a $2\times 2$ matrix, and it has been calculated at one loop in Refs. \citen{NKsingletopTev,NKtch} and at two loops in Ref. \citen{NKtch}. The elements of this matrix are given at one-loop by \cite{NKsingletopTev,NKtch}
\beqa
{\Gamma}_{S\, 11}^{t \, (1)}&=&
C_F \left[\ln\left(\frac{t(t-m_t^2)}{m_t s^{3/2}}\right)-\frac{1}{2}\right] 
\nonumber \\
{\Gamma}_{S\, 12}^{t \, (1)}&=&\frac{C_F}{2N} \ln\left(\frac{u(u-m_t^2)}{s(s-m_t^2)}\right)
\nonumber \\
{\Gamma}_{S\, 21}^{t \, (1)}&=& \ln\left(\frac{u(u-m_t^2)}{s(s-m_t^2)}\right)
\nonumber \\
{\Gamma}_{S\, 22}^{t \, (1)}&=& C_F \ln\left(\frac{s-m_t^2}{m_t \sqrt{s}}\right)
-\frac{1}{2N}\ln\left(\frac{t(t-m_t^2)}{s(s-m_t^2)}\right) 
+\frac{(N^2-2)}{2N}\ln\left(\frac{u(u-m_t^2)}{s(s-m_t^2)}\right)-\frac{C_F}{2}
\nonumber \\
\eeqa

At two loops, the first element of this matrix is given by \cite{NKtch}
\beq
\Gamma_{S\, 11}^{t \, (2)}=\frac{K}{2}\Gamma_{S\, 11}^{t \, (1)}
+C_F C_A \frac{(1-\zeta_3)}{4} \, .
\eeq

\begin{figure}
\centerline{\includegraphics[width=12cm]{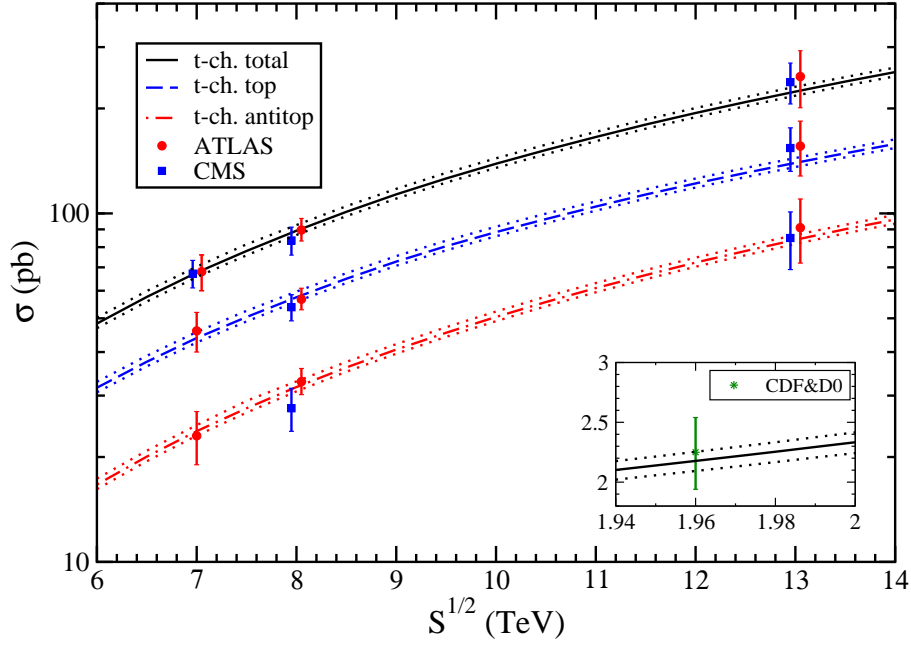}}
\caption{Single-top aNNLO production cross sections in the $t$-channel compared with (inset) CDF and D0 combination data at 1.96 TeV \cite{CDFD0tch}, and with ATLAS and CMS data at 7 TeV \cite{ATLAStch7,CMStch7}, 8 TeV \cite{ATLAStch8,CMStch8}, and 13 TeV \cite{ATLAStch13,CMStch13}.}
\label{tchplot}
\end{figure}

In Figure \ref{tchplot} we plot the aNNLO $t$-channel single-top and single-antitop cross sections, and their sum, with $m_t=172.5$ GeV with theoretical uncertainty from scale variation and the pdf error. Excellent agreement is found with D0 and CDF combination \cite{CDFD0tch} at 1.96 TeV energy, and with CMS \cite{CMStch7,CMStch8,CMStch13} and ATLAS \cite{ATLAStch7,ATLAStch8,ATLAStch13} results at 7, 8, and 13 TeV energies.  

The theoretical ratio $\sigma(t)/\sigma({\bar t})= 1.85{}^{+0.10}_{-0.08}$ at 
7 TeV compares well with the ATLAS \cite{ATLAStch7} result of $2.04 \pm 0.18$. The theoretical ratio $\sigma(t)/\sigma({\bar t})= 1.81{}^{+0.10}_{-0.07}$ at 8 TeV compares well with the ATLAS \cite{ATLAStch8} result of $1.72 \pm 0.09$ and the CMS \cite{CMStch8} result of $1.95 \pm 0.10 \pm 0.19$. The theoretical ratio for the total $t$-channel cross section $\sigma$(8 TeV)/$\sigma$(7 TeV)=$1.32{}^{+0.07}_{-0.05}$ compares well with the CMS \cite{CMStch8} result of $1.24 \pm 0.08 \pm 0.12$. The theoretical ratio $\sigma(t)/\sigma({\bar t})= 1.66{}^{+0.08}_{-0.06}$ at 13 TeV is in agreement with the ATLAS \cite{ATLAStch13} result of $1.72 \pm 0.09 \pm 0.18$ and the CMS result \cite{CMStch13} of $1.81 \pm 0.18 \pm 0.15$.

\begin{figure}
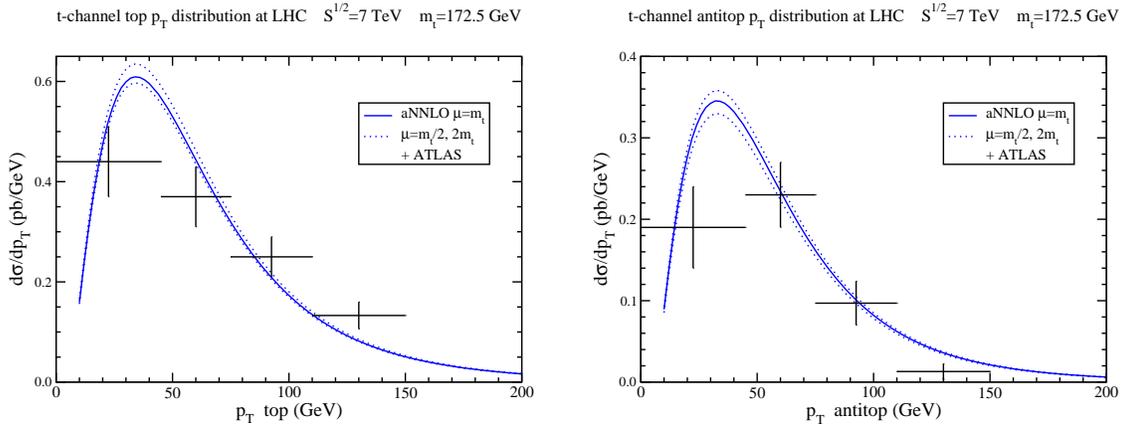

\centerline{\includegraphics[width=7cm]{pttoptchaNNLO7lhcATLAS1406.7844plot.eps}
\hspace{5mm}
\includegraphics[width=7cm]{ptantitoptchaNNLO7lhcATLAS1406.7844plot.eps}}
\caption{The aNNLO top-quark (left) and antitop (right) $p_T$ distributions in $t$-channel production at 7 TeV compared to ATLAS \cite{ATLAStch7} data.}
\label{pttch7}
\end{figure}

In addition to the total cross section, the top-quark $p_T$ distribution 
in $t$-channel production is of great interest. Figure \ref{pttch7} shows the top (left) and antitop (right) aNNLO $p_T$ distributions in $t$-channel production at 7 TeV LHC energy together with data from ATLAS \cite{ATLAStch7}.

\begin{figure}
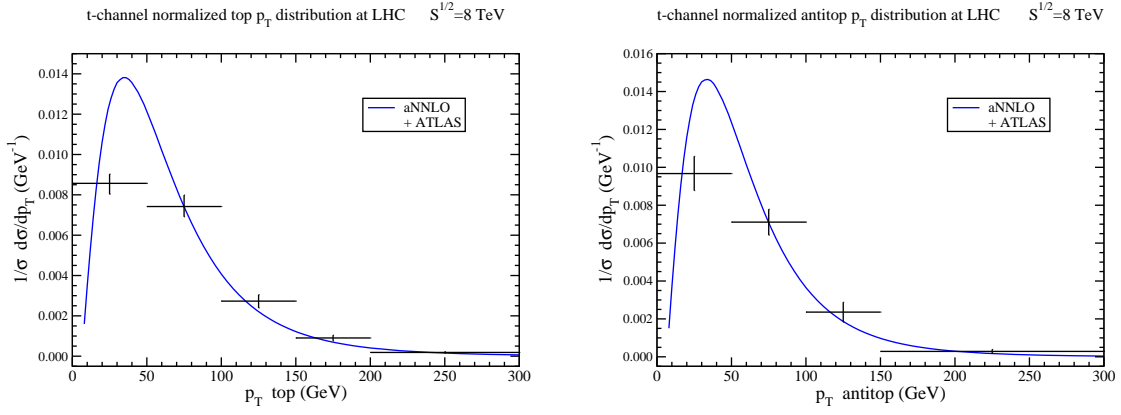

\centerline{\includegraphics[width=7cm]{ptnormtoptch8lhcATLAS1702.02859plot.eps}
\hspace{5mm}
\includegraphics[width=7cm]{ptnormantitoptch8lhcATLAS1702.02859plot.eps}}
\caption{The aNNLO top-quark (left) and antitop (right) normalized $p_T$ distributions in $t$-channel production at 8 TeV compared to ATLAS \cite{ATLAStch8} data.}
\label{pttch8}
\end{figure}

Figure \ref{pttch8} shows the top (left) and antitop (right) aNNLO normalized $p_T$ distributions in $t$-channel production at 8 TeV LHC energy. We find very good agreement between theory and data from ATLAS \cite{ATLAStch8} for both distributions.

\begin{figure}
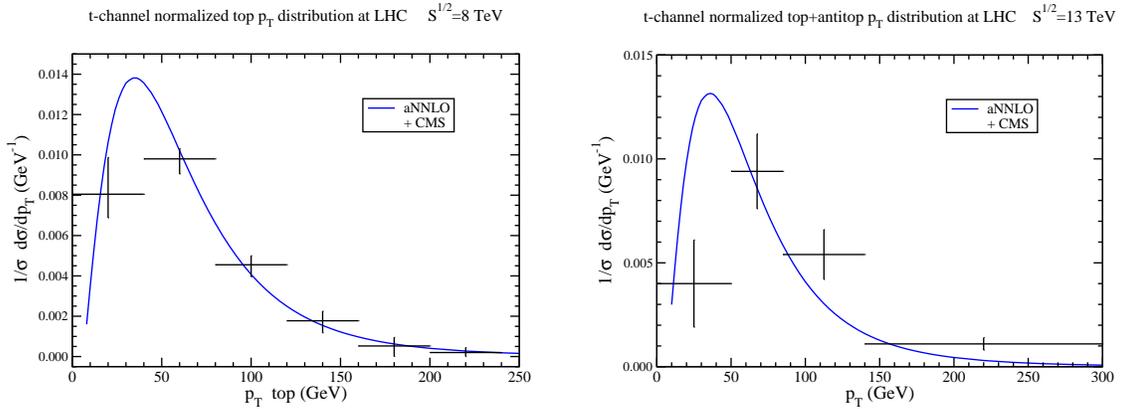

\centerline{\includegraphics[width=7cm]{ptnormtoptch8lhcCMS-TOP-14-004plot.eps}
\hspace{5mm}
\includegraphics[width=7cm]{ptnormtandtbartch13lhcCMS-TOP-16-004plot.eps}}
\caption{(Left) The aNNLO top-quark normalized $p_T$ distribution in $t$-channel production at 8 TeV compared to CMS \cite{CMStchpt8} data. (Right) The aNNLO top plus antitop normalized $p_T$ distribution in $t$-channel production at 13 TeV compared to CMS \cite{CMStchpt13} data.}
\label{pttch8and13}
\end{figure}

The left plot of Fig. \ref{pttch8and13} shows the top-quark aNNLO normalized $p_T$ distribution in $t$-channel production at 8 TeV LHC energy compared to CMS \cite{CMStchpt8} data. We note the very good agreement between theory and data. The plot on the right of Fig. \ref{pttch8and13} shows the top plus antitop aNNLO normalized $p_T$ distribution in $t$-channel production at 13 TeV LHC energy compared to CMS \cite{CMStchpt13} data.

\subsection{$s$-channel production}

We continue with single-top production in the $s$-channel.  

The soft anomalous dimension matrix for this process has been calculated at one loop\cite{NKsingletopTev,NKsch} and at two loops\cite{NKsch}. The $2 \times 2$ matrix for $s$-channel single-top production at one loop is \cite{NKsingletopTev,NKsch}
\beqa
\Gamma_{S\, 11}^{s \, (1)}&=&C_F \left[\ln\left(\frac{s-m_t^2}{m_t\sqrt{s}}\right)
-\frac{1}{2}\right]
\nonumber \\
\Gamma_{S\, 12}^{s \, (1)}&=&\frac{C_F}{2N} \ln\left(\frac{u(u-m_t^2)}{t(t-m_t^2)}\right)
\nonumber \\
\Gamma_{S\, 21}^{s \, (1)}&=& \ln\left(\frac{u(u-m_t^2)}{t(t-m_t^2)}\right)
\nonumber \\
\Gamma_{S\, 22}^{s \, (1)}&=&C_F \ln\left(\frac{s-m_t^2}{m_t \sqrt{s}}\right)
-\frac{1}{N}\ln\left(\frac{u(u-m_t^2)}{t(t-m_t^2)}\right)
+\frac{N}{2} \ln\left(\frac{u(u-m_t^2)}{s(s-m_t^2)}\right)-\frac{C_F}{2}
\nonumber \\
\eeqa
The first element of this $2 \times 2$ matrix at two loops is \cite{NKsch}
\beq
\Gamma_{S\, 11}^{s \, (2)}=\frac{K}{2} \Gamma_{S\, 11}^{s \, (1)}
+C_F C_A \frac{(1-\zeta_3)}{4} \, .
\eeq

\begin{figure}
\centerline{\includegraphics[width=11cm]{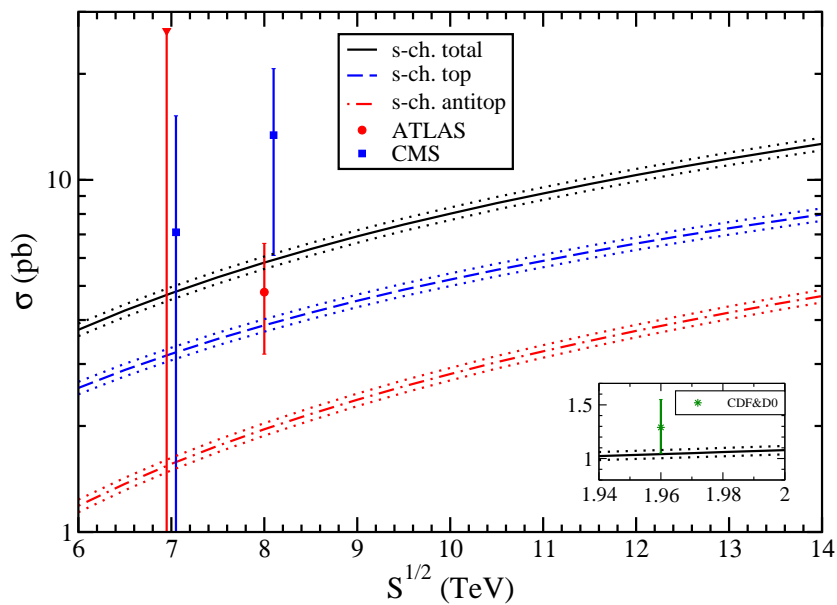}}
\caption{Single-top aNNLO production cross sections in the  $s$-channel compared to ATLAS and CMS data at 7 TeV \cite{ATLASsch7,CMSsch7and8} and 8 TeV \cite{CMSsch7and8,ATLASsch8}, and (inset) to CDF and D0 combination data \cite{CDFD0sch} at 1.96 TeV.}
\label{schlhcplot}
\end{figure}
Figure \ref{schlhcplot} shows the aNNLO cross sections for $s$-channel production with theoretical uncertainty from scale variation and pdf error. 
Results are shown for single-top production, single-antitop production, and their sum. Excellent agreement is found with 
D0 and CDF combination \cite{CDFD0sch}, CMS \cite{CMSsch7and8}, and ATLAS \cite{ATLASsch8} results.

\subsection{$tW$ production}

We continue with the associated production of a top quark with a $W$ boson.
The cross section for ${\bar t}W^+$ production is identical to that for $tW^-$.

The soft anomalous dimension for $bg \rightarrow tW^-$ is given at one loop by 
\cite{NKsingletopTev,NKtWH}
\beq
\Gamma_S^{tW \, (1)}=C_F \left[\ln\left(\frac{m_t^2-t}{m_t\sqrt{s}}\right)
-\frac{1}{2}\right] +\frac{C_A}{2} \ln\left(\frac{m_t^2-u}{m_t^2-t}\right)
\eeq
and at two loops by \cite{NKtWH}
\beq
\Gamma_S^{tW \, (2)}=\frac{K}{2} \Gamma_S^{tW \, (1)}
+C_F C_A \frac{(1-\zeta_3)}{4} \, .
\eeq

\begin{figure}
\centerline{\includegraphics[width=11cm]{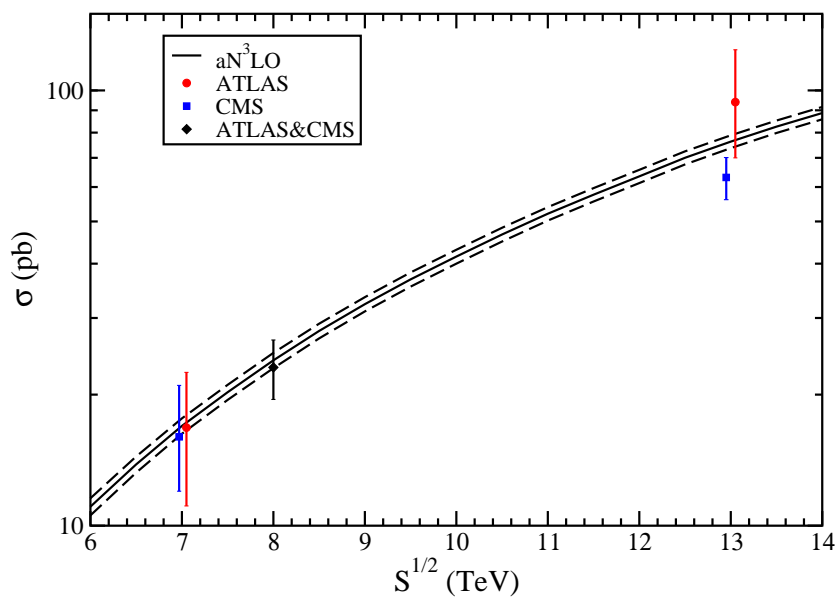}}
\caption{Single-top aN$^3$LO production cross sections for $tW$ production compared to ATLAS and CMS data at 7 TeV \cite{ATLAStW7,CMStW7}, 8 TeV \cite{ATLASCMStW8}, and 13 TeV \cite{ATLAStW13,CMStW13}.}
\label{tWlhcplot}
\end{figure}

Fig. \ref{tWlhcplot} shows the total $tW$ aN$^3$LO \cite{NKtW16} cross section (sum of $tW^-$ and ${\bar t} W^+$) as a function of LHC energy. Excellent agreement is found with data from ATLAS \cite{ATLAStW7} and CMS \cite{CMStW7} at 7 TeV, an ATLAS/CMS combination at 8 TeV \cite{ATLASCMStW8}, and ATLAS \cite{ATLAStW13} and CMS\cite{CMStW13} at 13 TeV. 

\begin{figure}
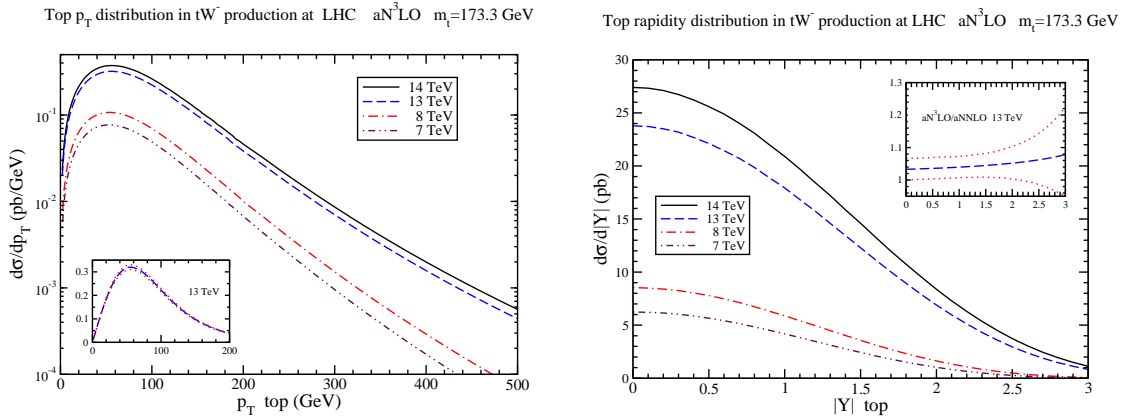

\centerline{\includegraphics[width=7cm]{pttoptWplot.eps}
\hspace{5mm} \includegraphics[width=7cm]{yabstoptWplot.eps}}
\caption{Top-quark $p_T$ (left) and rapidity (right) aN$^3$LO distributions in $tW^-$ production.}
\label{ptyabstW}
\end{figure}

Figure \ref{ptyabstW} displays the aN$^3$LO \cite{NKtW16} top-quark $p_T$ and rapidity distributions in $tW^-$ production at LHC energies.

\section{Top-quark production in models of new physics}

In addition to the various Standard-Model processes for top-quark production, some of which we studied above, other possibilities include top production in association with particles in models of new physics, or top production via top-quark anomalous couplings. We consider some of these possibilities below.

\subsection{Associated production of a top quark with a charged Higgs boson}

\begin{figure}
\centerline{\includegraphics[width=8cm]{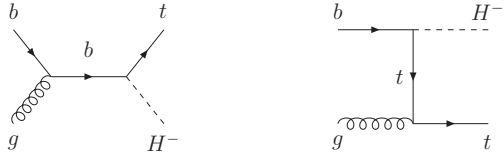}}
\caption{Lowest-order diagrams for the associated production of a top quark with a charged Higgs boson.}
\label{chiggsdiag}
\end{figure}

We first consider the production of a top quark in association with a charged Higgs boson \cite{NKcH,NKtWH,NKcH2}. Charged Higgs bosons appear in the Minimal Supersymmetric Standard Model (MSSM) and other two-Higgs doublet models.
The lowest-order diagrams for this process are shown in Fig. \ref{chiggsdiag}.
The soft anomalous dimension for this process is the same as for $tW$ production.

\begin{figure}
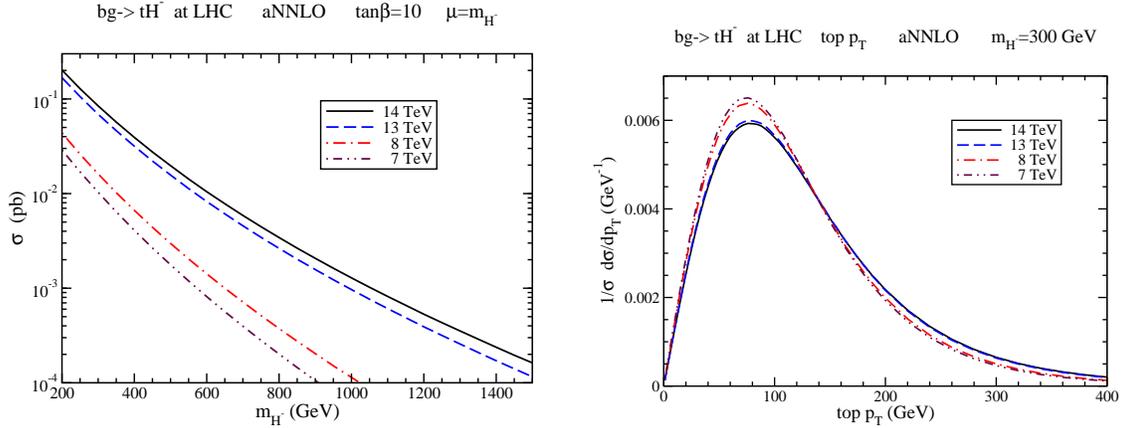

\centerline{\includegraphics[width=7cm]{chiggstn10plot.eps} \hspace{5mm}
\includegraphics[width=7cm]{ptnormtopchiggs300tn30plot.eps}}
\caption{Total cross sections (left) and normalized top-quark $p_T$ distributions (right) at aNNLO for charged-Higgs production in association with a top quark.}
\label{chiggslhcplot}
\end{figure}

The left plot in Fig. \ref{chiggslhcplot} shows the aNNLO cross section for $tH^-$ production in the MSSM, with $\tan\beta=10$, at LHC energies as a function of charged-Higgs mass. 
The aNNLO corrections increase the NLO cross section significantly, with the particular value of the increase depending on the charged-Higgs mass.
The plot on the right in Fig. \ref{chiggslhcplot} shows the aNNLO normalized top-quark $p_T$ distributions in $tH^-$ production for a charged-Higgs mass of 300 GeV.

\begin{figure}
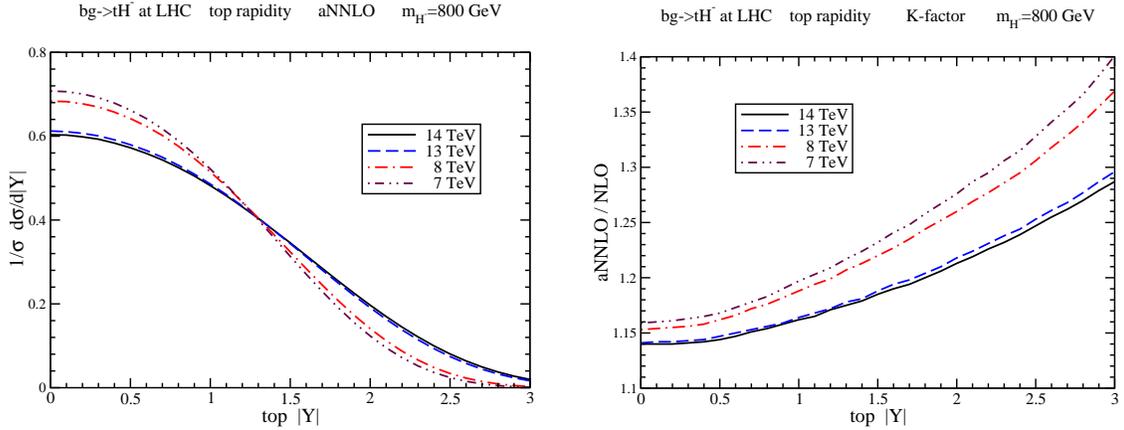

\centerline{\includegraphics[width=7cm]{yabsnormtopchiggs800tn30plot.eps}
\hspace{5mm}
\includegraphics[width=7cm]{Kyabstopchiggs800plot.eps}}
\caption{Normalized aNNLO top-quark rapidity distributions (left) and the corresponding $K$-factors (right) for charged-Higgs production in association with a top quark.}
\label{chiggsylhcplot}
\end{figure}

The left plot of Fig. \ref{chiggsylhcplot} shows the aNNLO top-quark rapidity distributions in $tH^-$ production for a charged-Higgs mass of 800 GeV at LHC energies. The plot on the right in Fig. \ref{chiggsylhcplot} shows the corresponding $K$ factors, which are very considerable, especially at large rapidity.

\subsection{Associated production of a top quark with a $Z$ boson via anomalous couplings}

\begin{figure}
\centerline{\includegraphics[width=8cm]{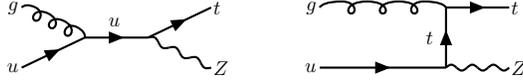}}
\caption{Lowest-order diagrams for the associated production of a top quark with a $Z$ boson via anomalous $t$-$q$-$Z$ coupling.}
\label{tZdiag}
\end{figure}

An interesting process that involves top-quark anomalous couplings is the associated production of a top quark with a $Z$ boson. While $tZ$ production can proceed via Standard Model processes involving an additional quark in the final state, it is possible to produce a $tZ$ final state without any other particles in models with anomalous couplings \cite{NKAB,NKtZ}. The lowest-order diagrams are shown in Fig. \ref{tZdiag}.

An effective Lagrangian that includes an anomalous coupling of a $t,q$ pair to a $Z$ boson is 
\beq
\Delta {\cal L}^{eff} =    \frac{1}{ \Lambda } \,
\kappa_{tqZ} \, e \, \bar t \, \sigma_{\mu\nu} \, q \, F^{\mu\nu}_Z + h.c.,
\eeq
where $\kappa_{tqZ}$ is the anomalous $t$-$q$-$Z$ coupling with 
$q$ an up or charm quark, 
$F^{\mu\nu}_Z$  is the $Z$-boson field tensor, 
$\sigma_{\mu \nu}=(i/2)(\gamma_{\mu}\gamma_{\nu}-\gamma_{\nu}\gamma_{\mu})$ 
with Dirac matrices $\gamma_{\mu}$,  and $\Lambda$ is an effective scale which 
we take to be the top-quark mass.

Soft-gluon corrections have been calculated at aNNLO in Ref. \cite{NKtZ}.
These soft corrections are important, and in fact at NLO they approximate the exact NLO \cite{NLOtqZ} results remarkably well for both $gu \rightarrow tZ$ and 
$gc \rightarrow tZ$. The aNNLO corrections provide additional enhancements. 

\begin{figure}
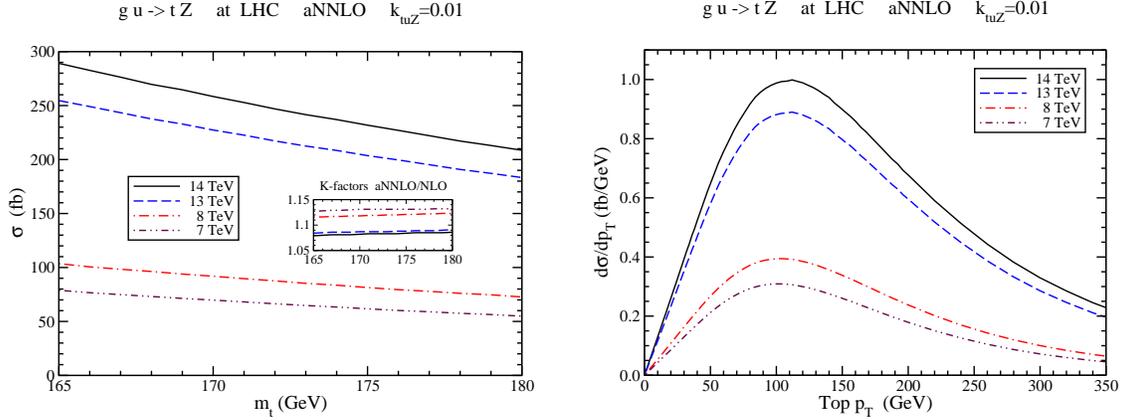

\centerline{\includegraphics[width=7cm]{gutZlhcplot.eps}
\hspace{5mm}
\includegraphics[width=7cm]{pttopgutZlhcplot.eps}}
\caption{Total cross sections (left) and top-quark $p_T$ distributions (right) at aNNLO for $tZ$ production via anomalous couplings.}
\label{tZlhcplot}
\end{figure}

The left plot of Fig. \ref{tZlhcplot} shows the aNNLO total cross sections for $gu\rightarrow tZ$ with $\kappa_{tuZ}=0.01$ at 7, 8, 13, and 14 TeV LHC energies as functions of the top-quark mass, with scales set equal to the top-quark mass. The inset plot shows the aNNLO/NLO $K$-factors at the various LHC energies. The aNNLO corrections increase the NLO result significantly at all LHC energies.

The right plot of Fig. \ref{tZlhcplot} shows the aNNLO top-quark $p_T$ distributions, $d\sigma/dp_T$, at 7, 8, 13, and 14 TeV LHC energies with $m_t=173.3$ GeV and  $\kappa_{tuZ}=0.01$. 

\subsection{Other top-quark production processes via anomalous couplings}

In addition to $tZ$ production discussed in the previous subsection, top quarks can also be produced in association with $Z'$ or $W'$ bosons \cite{MGNK} (see Ref. \cite{AFMG} for models of such particles). Soft-gluon corrections are significant for such processes \cite{MGNK}.

The production of top quarks with photons, $gq \rightarrow t \gamma$, via anomalous $t$-$q$-$\gamma$ couplings, with $q$ an up or charm quark, at Tevatron energy was studied in Ref. \cite{NKAB}. The soft-gluon corrections were found to be significant. The corrections are also large at LHC energies \cite{MFNK}.

The process $e u \rightarrow e t$ in electron-proton collisions via anomalous $t$-$u$-$\gamma$ coupling was studied in Ref. \cite{ABNK}. The processes $e q \rightarrow e t$, with $q$ an up or charm quark, via anomalous $t$-$q$-$\gamma$ and $t$-$q$-$Z$ couplings, were studied in Ref. \cite{NKAB}. 
The soft-gluon corrections were found to be important.

Same-sign top-quark production, $qq \rightarrow tt$, with $q$ an up or charm quark, via anomalous $t$-$q$-$\gamma$ and $t$-$q$-$Z$ couplings, was studied in  Ref. \cite{NKAB}. Numerical results were given for Tevatron energy \cite{NKAB}.

The process  $gu \rightarrow tg$ via anomalous $t$-$u$-$g$ couplings was studied in detail in Ref. \cite{NKEM}. The soft-gluon corrections were found to be substantial at LHC energies.

\section{Summary}

In this review, I have discussed soft-gluon corrections for top-quark 
production in hadronic collisions. I have presented the resummation of 
soft-gluon contributions in various top-quark processes through NNLL accuracy via two-loop calculations of soft anomalous dimension matrices.

N$^3$LO approximate results with soft-gluon corrections for the $t {\bar t}$ production cross section, and the top-quark differential distributions 
in transverse momentum and rapidity, are in excellent agreement 
with data from the LHC and the Tevatron.
Single-top cross sections and differential distributions have been presented in the $t$-channel and $s$-channel, and in $tW$ production, and they are also in excellent agreement with collider data.
 
Top-quark production in association with charged-Higgs bosons or with anomalous couplings in models of new physics has also been discussed.
Soft-gluon corrections are very significant for all top-quark production 
processes, and they reduce the theoretical errors.

\section*{Acknowledgments}

This material is based upon work supported by the National Science Foundation under Grant No. PHY 1519606.

\end{document}